\documentclass[12pt]{iopart}

\usepackage{bm}
\usepackage{color}
\usepackage{epsfig}
\usepackage{citesort}
\usepackage{graphicx}
\eqnobysec

\newcommand{\lwig}{\mbox{\;\raisebox{.3ex}
    {$<$}$\!\!\!\!\!$\raisebox{-.9ex}{$\sim$}\;}}
\newcommand{\gwig}{\mbox{\;\raisebox{.3ex}
    {$>$}$\!\!\!\!\!$\raisebox{-.9ex}{$\sim$}}\;}
\newcommand{\lambdabar}{{\hbox{$\lambda$\kern-1.ex\raise+0.45ex\hbox{--}}}}

\DeclareMathAlphabet{\mathpzc}{OT1}{pzc}{m}{it}

\begin{document}

\begin{flushright}
{\large \tt 
MPP-2008-111}
\end{flushright}

\title[Higher order corrections to the large scale matter power spectrum with neutrinos]%
{Higher order corrections to the large scale matter power spectrum in the presence of massive neutrinos}

\author{Yvonne~Y.~Y.~Wong\footnote{Present address:
Theory Division, Physics Department, CERN, CH-1211 Geneva 23, Switzerland}}
\address{Max-Planck-Institut f\"ur Physik (Werner-Heisenberg-Institut) \\
F\"ohringer Ring 6, D-80805 M\"unchen, Germany}

\ead{\mailto{ywong@mppmu.mpg.de}}

\begin{abstract}
We present the first systematic derivation of 
the one-loop correction to the large scale matter power spectrum in
a mixed cold+hot dark matter cosmology 
with subdominant massive neutrino hot dark matter.  Starting with the equations of motion for the 
density and velocity fields, we derive perturbative solutions to these quantities
and construct recursion relations for the interaction kernels, 
noting and justifying all approximations along the way. 
We find interaction kernels similar to those for a cold dark matter-only universe, 
but with additional 
dependences on the neutrino energy density fraction $f_\nu$ and the linear growth 
functions of the 
incoming wavevectors.  Compared with the $f_\nu=0$ case, 
the one-loop corrected matter power spectrum for a mixed
dark matter cosmology exhibits a decrease in small scale power
exceeding the canonical $\sim 8 f_\nu$ suppression predicted by linear theory,
a feature also seen in multi-component
$N$-body simulations.
\end{abstract}

\maketitle

\section{Introduction}

Standard big bang theory predicts a background of relic neutrinos, permeating the 
universe at an average of 112 neutrinos per cubic cm per neutrino flavour.
This enormous abundance means that even a sub-eV to eV neutrino
mass $m_\nu$  will render these otherwise elusive particles 
a significant dark matter component, $\Omega_\nu h^2 = \sum m_\nu/(93 \ {\rm eV})$.
Experimentally, a lower limit of
$\sum m_\nu \gwig 0.05 \ {\rm eV}$ on the sum of the neutrino masses 
has been established by neutrino oscillation experiments~(e.g., \cite{Amsler:2008zz}). 
On the other hand, tritium $\beta$-decay end-point
spectrum measurements point to an upper 
limit of $\sum m_\nu \lwig 6 \ {\rm eV}$~(e.g., \cite{Amsler:2008zz}). 
The corresponding energy densities, $0.001 \lwig \Omega_\nu \lwig 0.12$, make for
an interesting prediction to be tested against cosmological observations.

Importantly, neutrino dark matter is of the hot variety because of their inherent 
thermal velocity, which prevents them from clustering gravitationally
on scales smaller than their free-streaming length.  
This free-streaming effect then feedbacks into the evolution of the 
dominant cold dark matter (CDM) component 
via the gravitational source term, leading to a suppressed rate in the formation
of structures  on small length scales that is in 
principle manifest in the power spectrum of the large 
scale structure distribution~\cite{Bond:1980ha,Doroshkevich:1980zs,bib:khlopov,Shafi:1984ek,Schaefer:1989ua}.
At present, observations of the cosmic microwave background anisotropies,
galaxy clustering, and type Ia supernovae together engender a
conservative upper limit on the contribution of neutrino dark matter,
or equivalently, on the neutrino masses, of $\sum m_\nu \lwig 1 \ {\rm eV}$
within the $\Lambda$CDM framework.  See~\cite{Lesgourgues:2006nd,Hannestad:2006zg}
for recent reviews.

Many future cosmological probes will continue to improve on this limit, or perhaps 
even detect neutrino dark matter (e.g., \cite{Lesgourgues:2006nd,Abdalla:2007ut,%
Hannestad:2007cp,Hannestad:2006as,%
Kitching:2008dp,Lesgourgues:2005yv,Gratton:2007tb,Ichikawa:2005hi,%
Lesgourgues:2007ix,Pritchard:2008wy}).  
To this end, it is important to note that many of these probes,
particularly high redshift galaxy surveys and 
weak gravitational lensing, 
will derive most of their constraining power at wavenumbers
$k \sim 0.1 \to 1 \ h \ {\rm Mpc}^{-1}$,
where the evolution of density perturbations has become weakly nonlinear.
Coincidentally, these are also the scales 
at which neutrino free-streaming is expected to produce the largest effect
on the large scale matter power spectrum. 
Thus, it is crucial that we understand the nonlinear evolution of density perturbations 
at these scales, in order to maximise our gain in detecting/constraining neutrino dark matter.

A number of recent works have attempted to model the effects of massive neutrinos on
the matter power spectrum at nonlinear scales using various different techniques.
These include multi-component $N$-body simulations~\cite{Brandbyge:2008rv} and semi-analytic
halo models~\cite{Abazajian:2004zh,Hannestad:2005bt,Hannestad:2006as}. 
However, with the exception of $N$-body simulations, these methods 
all contain elements that require prior calibration against simulation results, and
not surprisingly, all calibrations to date have been performed in a $\Lambda$CDM setting.  
This renders these nonlinear models at present not completely satisfactory for
use with cosmologies containing massive neutrinos.
Recalibration against simulations in the appropriate cosmology will, however, enhance/restore our 
confidence in these methods.

Recently, Saito {\it et al.}~\cite{Saito:2008bp} proposed an alternative method 
based on higher order cosmological perturbation theory.  Since its inception in the 
1980s~\cite{bib:juszkie,bib:vishniac,Fry:1983cj,Goroff:1986ep},
higher order cosmological perturbation theory has found numerous applications ranging
from computation of the weakly nonlinear power spectrum and gravity-induced 
bispectrum, to the exploration of  nonlinear galaxy bias.
  See reference~\cite{Bernardeau:2001qr} for a review.
In the perturbative  approach, one envisages nonlinear evolution as the outcome of 
interactions of a collection of linear waves (perturbations).  
The equations of motion of the system define the interaction kernels.

Saito {\it et al.}'s recipe is simple:  assume the neutrino density perturbations  remain linear
at all times, and apply nonlinear modelling only to the CDM+baryon component.  This basic scheme 
has also been adopted in some earlier nonlinear models including 
 massive neutrinos~\cite{Hannestad:2006as}.  
For $\sum m_\nu \lwig 0.6 \ {\rm eV}$, multi-component
$N$-body simulations have confirmed that a linear approximation 
for the neutrino density contrast is sound~\cite{Brandbyge:2008rv}.

For the CDM+baryon component, Saito {\it et al.}~calculated the one-loop correction
to the CDM+baryon auto-correlation power spectrum, using
the correctly computed linear waves (perturbations), but interaction kernels that 
have been developed for a CDM-only universe.
This amounts to ignoring additional mode-coupling effects between the linear growth functions at different
wavenumbers.  Although, as we shall show, this approximation does lead to a considerable
simplification in the final form of the nonlinear  power spectrum, it is nonetheless
reminiscent of the mismatch between the cosmology to which we apply
and that on which we calibrate some of the semi-analytic methods discussed above, and therefore calls for
closer scrutiny.

In this paper we present a systematic derivation of the one-loop correction
to the matter power spectrum in the presence of massive neutrinos from first principles.
As in~\cite{Hannestad:2006as,Saito:2008bp},
we assume linearity for the neutrino component.
For CDM+baryons, however, we begin with the relevant equations of motion,
and solve them (approximately) perturbatively in an Einstein--de Sitter $\Omega_m=1$ 
background at high redshifts ($z \gwig 0.5$).
From these solutions we derive interaction kernels that capture also the physics of 
mode-coupling between linear growth functions at different 
wavenumbers.  With these kernels we compute the correct nonlinear power spectrum.

In section~\ref{sec:eom} we give the relevant equations of motion.  We discuss  briefly
the linear order solutions in section~\ref{sec:linear}, before formulating our higher order
perturbative approach in section~\ref{sec:beyond}.  In section~\ref{sec:approx} we outline
our scheme to obtain approximate solutions to the equations of motion for higher order
perturbations, while in section~\ref{sec:nthorder} we derive recursion relations
for the interaction kernels 
and thus generalise our approximate solutions to arbitrary orders in perturbative expansion.
Section~\ref{sec:spectra} deals with the calculation of the one-loop correction to
the matter power spectrum, which we evaluate numerically for realistic cosmological models
and discuss in detail  in section~\ref{sec:discuss}.
We conclude in section~\ref{sec:conclusions}.

\section{Equations of motion\label{sec:eom}}

We begin with the standard set of equations of motion for the density perturbations $\delta ({\bm x},\tau)$ 
and the peculiar velocity ${\bm u}({\bm x},\tau)$ of the CDM+baryon component (e.g., \cite{Bernardeau:2001qr}),
\begin{eqnarray}
\label{eq:eomx}
\frac{\partial \delta ({\bm x},\tau)}{\partial \tau} + \nabla \cdot \{ [1+\delta({\bm x},\tau)] {\bm u} ({\bm x},\tau) \}=0, \nonumber \\
\frac{\partial {\bm u}({\bm x},\tau)}{\partial \tau} + {\cal H}(\tau) {\bm u}({\bm x},\tau) + [{\bm u}({\bm x},\tau) \cdot \nabla] {\bm u}
({\bm x},\tau)+\nabla \Phi({\bm x},\tau)=0,
\end{eqnarray}
where $\tau$ is the conformal time, ${\bm x}$ the comoving coordinates, 
${\cal H} \equiv d \ln a/d \tau = Ha$  the conformal expansion rate, and
the stress tensor has been set explicitly to zero.
The Poisson equation
\begin{equation}
\label{eq:poisson}
\nabla^2 \Phi({\bm x},\tau) = \frac{3}{2} {\cal H}^2(\tau) \Omega_m(\tau)  [(1-f_\nu) \delta({\bm x},\tau) + f_\nu
 \delta^\nu({\bm x},\tau)]
\end{equation}
relates the Newtonian gravitational potential $\Phi({\bm x},\tau)$ to the
density perturbations.  Perturbations from both CDM+baryons $\delta({\bm x},\tau)$ 
and neutrinos $\delta^\nu({\bm x},\tau)$ 
contribute to this gravitational source term, although in the latter case
the contribution is suppressed by the neutrino density fraction
$f_\nu \equiv \Omega_\nu(\tau)/\Omega_m(\tau)=\Omega_{\nu,0}/\Omega_{m,0}$.  This 
fraction is constant in time once the neutrinos have become nonrelativistic.

It is common to rewrite the equations of motion in Fourier space.
Defining the Fourier transform
\begin{equation}
\tilde{\phi}({\bm k},\tau) = \int \frac{d^3{\bm x}}{(2 \pi)^3} \exp(-i {\bm k} \cdot {\bm x}) \phi({\bm x},\tau)
\end{equation}
for some field $\phi({\bm x},\tau)$,
and the divergence of the velocity field
\begin{equation}
\theta({\bm x},\tau) \equiv \nabla \cdot {\bm u}({\bm x},\tau),
\end{equation}
equations~(\ref{eq:eomx}) and~(\ref{eq:poisson}) can be equivalently expressed as
\begin{eqnarray}
\label{eq:eomk}
 \frac{\partial \tilde{\delta}({\bm k},\tau)}{\partial \tau} + \tilde{\theta} ({\bm k},\tau) = 
-\int d^3 {\bm q}_1 d^3 {\bm q}_2 \delta_D({\bm k}-{\bm q}_{12}) \alpha({\bm q}_1,{\bm q}_2)
\tilde{\theta}({\bm q}_1,\tau) \tilde{\delta}({\bm q}_2,\tau),
\nonumber \\
 \frac{\partial \tilde{\theta}({\bm k},\tau)}{\partial \tau} + 
 {\cal H}(\tau)\tilde{\theta}({\bm k},\tau) + \frac{3}{2} {\cal H}^2(\tau)
\Omega_m(\tau)
[(1-f_\nu) \tilde{\delta}({\bm k},\tau)+f_\nu \tilde{\delta}^\nu({\bm k},\tau)] = \nonumber \\
\hspace{25mm} 
- \int d^3{\bm q}_1 d^3 {\bm q}_2 \delta_D({\bm k}-{\bm q}_{12}) \beta({\bm q}_1,{\bm q}_2) 
\tilde{\theta}({\bm q}_1,\tau) \tilde{\theta}({\bm q}_2,\tau),
\end{eqnarray}
where
\begin{equation}
\alpha({\bm q}_1,{\bm q}_2) \equiv \frac{{\bm q}_{12} \cdot {\bm q}_1}{q_1^2}, 
\qquad \beta({\bm q}_1,{\bm q}_2) \equiv \frac{q_{12}^2 ({\bm q}_1 \cdot
{\bm q}_2)}{2 q_1^2 q_2^2}.
\end{equation}
Here, we have defined $q\equiv |{\bm q}|$, and ${\bm q}_{i \cdots j}\equiv {\bm q}_i+\cdots +{\bm q}_j$.  
The function 
$\delta_D$ denotes the Dirac delta function.
As usual, we assume the vorticity of the velocity field 
${\bm w}({\bm x},\tau) \equiv \nabla \times {\bm u}({\bm x},\tau)$ to be
negligible at all times.

For the neutrino component, a full treatment requires that we track the evolution of 
the neutrino phase space density ${\mathpzc f}^\nu({\bm x},{\bm p},\tau)$ by
solving the Vlasov equation (e.g., \cite{Bernardeau:2001qr}),
\begin{equation}
\label{eq:vlasov}
\frac{\partial {\mathpzc f}^\nu}{\partial \tau} + \frac{\bm p}{m_\nu a} \cdot 
\nabla {\mathpzc f}^\nu - a m_\nu \nabla \Phi \cdot \frac{\partial {\mathpzc f}^\nu}{\partial {\bm p}}=0,
\end{equation}
where $m_\nu$ is the neutrino mass, and we have assumed the neutrinos to be nonrelativistic.
Integrating over the neutrino momentum ${\bm p}$,
\begin{equation}
\label{eq:phasespace}
\int d^3{\bm p} \ {\mathpzc f}^\nu({\bm x},{\bm p},\tau) \equiv \bar{\rho}^\nu(\tau)[1+\delta^\nu({\bm x},\tau)],
\end{equation}
yields the neutrino density contrast $\delta^\nu({\bm x},\tau)$.

\section{Linear theory\label{sec:linear}}

Linearising the equations of motion~(\ref{eq:eomk}),
i.e., dropping all terms on the r.h.s.\ of the equal sign, 
the growing mode 
solutions for the CDM+baryon density  contrast and peculiar velocity  at 
some wavevector ${\bm k}$ can be written as
\begin{eqnarray}
\delta_1({\bm k},\tau)=D_1(k,\tau) \delta({\bm k},\tau_0),  \nonumber \\
\theta_1({\bm k},\tau)= -{\cal H}(\tau) f(k,\tau) D_1(k,\tau) \delta({\bm k},\tau_0),
\end{eqnarray}
where $\tau_0$ denotes some initial time well in the matter-domination epoch,
$D_1(k,\tau)$ is the linear growth function, its logarithmic derivative
\begin{equation}
\label{eq:ffunc}
f(k,\tau) \equiv \frac{\partial \ln D_1 (k,\tau)}{\partial \ln a} = \frac{1}{\cal H} 
\frac{\partial \ln D_1(k,\tau)}{\partial \tau},
\end{equation}
and we have dropped the tildes on $\tilde{\delta}_1({\bm k},\tau)$ and $\tilde{\theta}_1({\bm k},\tau)$ for convenience.
A similar expression,
\begin{equation}
\delta^\nu_1({\bm k},\tau)=D^\nu_1(k,\tau) \delta^\nu({\bm k},\tau_0),
\end{equation}
describes the growth of the neutrino density perturbations in the linear regime.

The inclusion of massive neutrinos in the matter content of the universe introduces 
a new length scale to the problem, the neutrino free-streaming scale $k_{\rm FS}$
(e.g., \cite{Ringwald:2004np}),
\begin{eqnarray}
k_{\rm FS}(\tau) \equiv \sqrt{ 
\frac{3 \Omega_m(\tau) {\cal H}^2(\tau)}{2 c^2_\nu }}
\simeq  1.5 \sqrt{a(\tau) \Omega_{m,0}} \left(\frac{m_\nu}{\rm eV} \right) \ h \ {\rm Mpc}^{-1},
\end{eqnarray}
where
\begin{equation}
c_\nu \equiv \frac{T_\nu(\tau)}{m_\nu} \sqrt{\frac{3 \zeta(3)}{2 \ln(2)}} \simeq \frac{81}{a(\tau)} 
\left(\frac{\rm eV}{m_\nu} \right) \ {\rm km \ s}^{-1}.
\end{equation}
At wavenumbers $k \ll k_{\rm FS}$, neutrinos cluster gravitationally and 
behave essentially like CDM.   In the other limit $k \gg k_{\rm FS}$, the neutrinos'
inherent thermal velocity $c_\nu$ prevents efficient infall into gravitational potential wells,
thereby suppressing the neutrino density perturbations 
relatively to their CDM+baryon counterparts. 

This qualitative picture is generally true for any cosmology.  However,
if we restrict our considerations to an Einstein--de Sitter universe, then using the 
relations $a \propto \tau^2$ and ${\cal H}(\tau) = 2/\tau$ it is easy to show that
\begin{eqnarray}
\label{eq:d1}
D_1(k,\tau) \propto a, &\qquad  f(k,\tau) = 1, &\qquad k \ll k_{\rm FS}, \nonumber \\
D_1(k,\tau) \propto  a^{1-\mu}, &\qquad  f(k,\tau) = 1-\mu , &\qquad  k \gg k_{\rm FS},
\end{eqnarray}
with
\begin{equation}
\mu=\frac{5}{4}-\frac{\sqrt{1+24(1-f_\nu)}}{4} \simeq \frac{3}{5} f_\nu.
\end{equation}
The $k \gg k_{\rm FS}$ solution is obtained by  
setting $\delta^\nu({\bm k},\tau)$ explicitly to zero in 
the equations of motion. For the neutrino component, the expressions
\begin{eqnarray}
D_1^\nu(k,\tau) \simeq D_1(k,\tau) \frac{\delta (k,\tau_0)}{\delta^\nu (k,\tau_0)}, &\qquad k \ll k_{\rm FS}, \nonumber \\
D_1^\nu(k,\tau) \simeq  D_1(k,\tau) 
\frac{\delta (k,\tau_0)}{\delta^\nu (k,\tau_0)} \left[\frac{k_{FS}^2(\tau) (1-f_\nu)}{k^2-k_{\rm FS}^2(\tau) f_\nu}\right],
&\qquad  k \gg k_{\rm FS}
\end{eqnarray}
have been shown to be asymptotic solutions to equation~(\ref{eq:vlasov})~\cite{Ringwald:2004np}.
For intermediate $k$ values, the linear growth functions for both the CDM+baryon and the neutrino 
components must be calculated numerically with a Boltzmann code such as {\tt CAMB}~\cite{Lewis:1999bs},
which also gives the neutrino sector a full general relativistic treatment~\cite{Ma:1995ey}.

\subsection{Fitting formulae}

We shall need estimates of the linear growth function and particularly 
its logarithmic derivative later in the analysis.   For this we  resort to fitting formulae.

For the CDM+baryon component, we have~\cite{Hu:1997vi}
\begin{equation}
\label{eq:fitting}
D_1(k,\tau) \simeq \left[ 1+ r^c(k,\tau) \right]^{\mu/c}
D^{1-\mu}_1(\tau),
\end{equation}
and hence
\begin{equation}
\label{eq:f}
f(k,\tau) \simeq 1 -\frac{\mu}{1+r^c(k,\tau)},
\end{equation}
where $c \simeq 0.7$, and 
\begin{eqnarray}
r(k,\tau) \equiv \frac{D_1(\tau)}{1+y_{\rm FS}(k)} , \nonumber \\
y_{\rm FS}(k) = 17.2 f_\nu (1+ 0.488 f_\nu^{-7/6}) (p N_\nu/f_\nu)^2, \nonumber \\
p = \left(\frac{k}{{\rm Mpc}^{-1}}\right) \Theta_{2.7}^2 (\Omega_{m,0} h^2)^{-1}.
\end{eqnarray}
Here, $D_1(\tau)$ is the growth function in the absence of massive neutrinos, 
normalised such that $D_1(\tau)= a/a_{\rm eq}$ in an Einstein--de Sitter universe.
The quantity $N_\nu$ is the number of massive neutrinos (assuming equal masses) which we generally 
take to be three, and
$\Theta_{2.7}$ is defined through $T_{\rm CMB}= 2.7 \ \Theta_{2.7} \ {\rm K}$. 
The claimed accuracy of the formula~(\ref{eq:fitting}) 
 is 1 to 2 \%~\cite{Hu:1997vi}.

For the neutrino component, we find 
\begin{equation}
\label{eq:nufitting}
D_1^\nu(k,\tau) \simeq  D_1(k,\tau) 
\frac{\delta (k,\tau_0)}{\delta^\nu (k,\tau_0)} \left[\frac{k_{\rm FS}^2(\tau) 
(1-f_\nu)}{[k+k_{\rm FS}(\tau)]^2-k_{\rm FS}^2(\tau) f_\nu}\right]
\end{equation}
to be a good interpolation between the $k \ll k_{\rm FS}$ and $k \gg k_{\rm FS}$ limits, accurate to
better than 5 \%.

\subsection{Linear power spectrum}

The linear matter power spectrum is defined as
\begin{equation}
\label{eq:linpow}
P^L(k,\tau) \delta_D({\bm k}+{\bm k}') \equiv \langle \delta^T_1({\bm k},\tau) \delta^T_1({\bm k}',\tau) \rangle,
\end{equation}
where $\delta^T_1({\bm k},\tau)\equiv f_{cb} \delta_1({\bm k},\tau)+ f_\nu \delta^\nu_1({\bm k},\tau)$,
with $f_{cb}=(\Omega_c+\Omega_b)/\Omega_m$,
counts both the CDM+baryon and the neutrino density contrast.
In terms of the linear growth functions,
\begin{equation}
P^L(k,\tau) 
= f^2_{cb} P^L_{cb}(k,\tau) + 2 f_{cb} f_\nu P^L_{cb \nu} (k,\tau) + f_\nu^2 P^L_{\nu}(k, \tau),
\end{equation}
with
\begin{eqnarray}
\label{eq:pcomponents}
P_{cb}^L(k,\tau)&=& [D_1(k,\tau)T_{cb}(k,\tau_0)]^2 P^I(k), \nonumber \\
P_{\nu}^L(k,\tau)&=& [D^\nu_1(k,\tau) T_\nu (k,\tau_0)]^2 P^I(k,), \nonumber \\
P_{cb\nu}^L(k,\tau)&=& D_1(k,\tau) D^\nu_1(k,\tau) T_{cb}(k,\tau_0) T_\nu (k,\tau_0)P^I(k),
\end{eqnarray}
where $T_i(k,\tau_0)$ are the linear transfer functions mapping the initial fluctuations
(from, e.g., inflation) through the epochs of horizon crossing and matter--radiation equality
to time $\tau_0$, i.e.,
$\delta({\bm k},\tau_0) = T(k,\tau_0) \delta^I({\bm k})$.
We have also assumed in equation~(\ref{eq:pcomponents}) adiabatic initial conditions 
so that $P^I_{cb}(k)
=P^I_{\nu}(k)=P^I_{cb\nu}(k)\equiv P^I(k)$.
Note that the power spectrum defined in this manner contributes 
$4 \pi k^3 P(k)$ per logarithmic wavenumber to the variance, in contrast to, e.g., 
the default output of {\tt CAMB}, which
contributes $k^3 P(k)/(2 \pi^2)$ between $\ln k$ and  $\ln k + d \ln k$.

\section{Beyond linear theory\label{sec:beyond}}

We wish to find a higher order perturbative description for the CDM+baryon density contrast
and peculiar velocity.  
To do so it is convenient to define a new time variable
\begin{equation}
s \equiv \ln a(\tau)= {\cal H}(\tau)  d\tau,
\end{equation}
and the vectors
\begin{equation}
\Psi({\bm k},s) \equiv \left[ \begin{array}{c}
				\delta({\bm k},s) \\
				-\frac{1}{{\cal H}(s)} \theta({\bm k},s)\\
				\end{array} \right], \qquad
\Psi^\nu({\bm k},s) \equiv \left[ \begin{array}{c}
				\delta^\nu({\bm k},s) \\
				0 \\
				\end{array} \right].
\end{equation}
The equations of motion~(\ref{eq:eomk}) can then be written in a more compact form,
\begin{eqnarray}
\label{eq:eomvector}
 \partial_s \Psi_a({\bm k},s) + K_{ab} \Psi_b({\bm k},s) 
+ N_{ab} \Psi^\nu_b({\bm k},s)
= \nonumber \\
\hspace{20mm} 
\int d^3{\bm q}_1 d^3 {\bm q}_2 \ \delta_D({\bm k}-{\bm q}_{12})
\gamma_{abc} ({\bm q}_1,{\bm q}_2) \Psi_b({\bm q}_1,s) \Psi_c({\bm q}_2,s),
\end{eqnarray}
where $a,b,c=1,2$, and repeated indices imply summation.

In the homogeneous part of equation~(\ref{eq:eomvector}), the matrix $K$ is given by
\begin{equation}
K =  \left[ \begin{array}{cc}
		0 & -1 \\
		- \frac{3}{2} (1-f_\nu) & \frac{1}{2} \end{array} \right],
\end{equation}
where we have assumed explicitly an Einstein--de Sitter universe.
The inhomogeneous part is specified by the matrix,
\begin{equation}
N = \left[ \begin{array}{cc}
		0 & 0 \\
		- \frac{3}{2} f_\nu & 0 \end{array} \right],
\end{equation}
and the tensor $\gamma_{abc} ({\bm q}_1,{\bm q}_2)$ is zero  except for
$\gamma_{121}=\alpha({\bm q}_1,{\bm q}_2)$ and 
$\gamma_{222}=\beta({\bm q}_1,{\bm q}_2)$.

We seek a perturbative solution of the form
\begin{equation}
\Psi({\bm k},s) = \sum_{n=1}^{\infty} \psi^{(n)}({\bm k},s).
\end{equation}
For the neutrinos, we assume only the linear solution is nonzero, i.e.,
\begin{equation}
\Psi^{\nu}({\bm k},s)=\left[ \begin{array}{c}
				\delta^\nu_1({\bm k},s) \\
				0 \\
				\end{array} \right].
\end{equation}
Then the equation of motion for the $n$th order solution, where $n \geq 2$, is
\begin{equation}
\label{eq:norder}
\partial_s \psi^{(n)}_a({\bm k},s) + K_{ab} \psi^{(n)}_b({\bm k},s) 
= B_a^{(n)}({\bm k},s),
\end{equation}
with
\begin{eqnarray}
B_a^{(n)}({\bm k},s)  =   \int 
d^3{\bm q}_1 d^3 {\bm q}_2 \delta_D({\bm k}-{\bm q}_{12})
\gamma_{abc} ({\bm q}_1,{\bm q}_2) 
\nonumber \\
\hspace{60mm} 
\times \sum_{m=1}^{n-1} \psi_b^{(n-m)}({\bm q}_1,s) \psi_c^{(m)}({\bm q}_2,s).
\end{eqnarray}
Observe how the term proportional to $\Psi^\nu({\bm k},s)$ has disappeared 
for $n>1$ because of the assumption that 
neutrino density perturbations remain linear at all times.

\section{Approximate solutions\label{sec:approx}}

Equation~(\ref{eq:norder}) can be solved by first defining a transformation matrix $U$ via
\begin{equation}
\tilde{K} \equiv U^{-1} K U = {\rm Diag}(\kappa_1,\kappa_2),
\end{equation}
so that in the diagonal basis, the equation of motion becomes
\begin{equation}
\label{eq:eom_transformed}
\partial_s \tilde{\psi}^{(n)}_a({\bm k},s) = -  \tilde{K}_{ab} \tilde{\psi}^{(n)}_b({\bm k},s) 
+ \tilde{B}^{(n)}_a({\bm k},s),
\end{equation}
where $\tilde{\psi}^{(n)}=U^{-1} \psi^{(n)}$, and $\tilde{B}^{(n)}=U^{-1} B^{(n)}$.
The formal solution to equation~(\ref{eq:eom_transformed}) is simple,
\begin{eqnarray}
\label{eq:exact}
\tilde{\psi}^{(n)}_a({\bm k},s) &=&  e^{-\int_{s_0}^{s} \kappa_a ds'} \tilde{\psi}^{(n)}_a({\bm k},s_0) 
+\int_{s_0}^s  e^{-\int_{s'}^{s} \kappa_a ds''} \tilde{B}^{(n)}_a({\bm k},s') ds' \nonumber \\
&=& \int_{s_0}^s  e^{-\int_{s'}^{s} \kappa_a ds''} \tilde{B}^{(n)}_a({\bm k},s') ds',
\end{eqnarray}
where we have used $\tilde{\psi}^{(n)}_a({\bm k},s_0) =0$ in the second line, 
i.e., the perturbations at the initial time $s_0$ are completely described by linear theory.
Within our formulation of the problem  the solution~(\ref{eq:exact}) is exact. 
To gain further ground, however, we must make some approximations, as we show order by order 
below.

(i) The $n=1$ solution is by definition
\begin{equation}
\psi^{(1)}_a({\bm k},s) = \int  d^3{\bm q} \delta_D({\bm k}-{\bm q}) Q^{(1)}_{a}({\bm q},s) 
\delta_1({\bm q},s),
\end{equation}
where $Q^{(1)}_1({\bm q},s) =1$, and $Q^{(1)}_2({\bm q},s) =f(q,s)$.
 
(ii) For $n=2$, we have
\begin{eqnarray}
\tilde{B}^{(2)}_a({\bm k},s) \!&=& \! \! \int d^3 {\bm q}_1 d^3 {\bm q}_2 \delta_D({\bm k}-{\bm q}_{12})
D_1(q_1,s) D_1(q_2,s) 
\nonumber \\
&& \!\!
\times    U^{-1}_{ab}  \gamma_{bcd} ({\bm q}_1,{\bm q}_2) Q_c^{(1)}({\bm q}_1,s)  Q_d^{(1)}({\bm q}_2,s)  
\delta({\bm q}_1,s_0) \delta({\bm q}_2,s_0).
\end{eqnarray}
The solution in the diagonal basis is then 
\begin{eqnarray}
\tilde{\psi}^{(2)}_a({\bm k},s)= \int_{s_0}^s  ds'
\int d^3{\bm q}_1 d^3 {\bm q}_2 \
 e^{-\int_{s'}^{s} \kappa_a ds''} e^{[h(q_1,s')+h(q_2,s')] s'}  \nonumber \\
\hspace{35mm} \times \delta_D({\bm k}-{\bm q}_{12})  
 U^{-1}_{ab}\gamma_{bcd} ({\bm q}_1,{\bm q}_2) \nonumber \\
\hspace{40mm}	\times		Q_c^{(1)}({\bm q}_1,s')  Q_d^{(1)}({\bm q}_2,s')  \delta({\bm q}_1,s_0) \delta({\bm q}_2,s_0),
\end{eqnarray}
where we have defined $h(q,s) \equiv \ln D_1(q,s)/\ln a(s)$.

The integration over $s'$ can be performed using integration by parts, i.e.,
\begin{eqnarray}
\label{eq:byparts}
\int_{s_0}^s  ds' Q_c^{(1)}({\bm q}_1,s')  Q_d^{(1)}({\bm q}_2,s')
e^{-\int_{s'}^{s} \kappa_a ds''} e^{[h(q_1,s')+h(q_2,s')] s'} =\nonumber \\
 \hspace{5mm} \left. \frac{ Q_c^{(1)}({\bm q}_1,s')  Q_d^{(1)}({\bm q}_2,s')}%
{\kappa_a + f(q_1,s')+f(q_2,s')} 
e^{-\int_{s'}^{s} \kappa_a ds''}e^{[h(q_1,s')+h(q_2,s')] s'}
\right|^s_{s_0} \nonumber \\
\hspace{10mm}  - \int_{s_0}^s  ds' Q_c^{(1)}({\bm q}_1,s') 
 Q_d^{(1)}({\bm q}_2,s') e^{-\int_{s'}^{s} \kappa_a ds''} e^{[h(q_1,s')+h(q_2,s')] s'} \nonumber \\
\hspace{18mm} \times  \frac{1}{Q_c^{(1)}({\bm q}_1,s')  Q_d^{(1)}({\bm q}_2,s')} \frac{\partial }{\partial s'} 
\left[\frac{Q_c^{(1)}({\bm q}_1,s')  Q_d^{(1)}({\bm q}_2,s')}{\kappa_a + f(q_1,s')+f(q_2,s')}\right],
\end{eqnarray}
and we have used the relation $h(q,s)+(\partial h/\partial s) s = f(q,s)$.
Compared with the original integrand on the l.h.s., the integrand
 on the r.h.s.\ of equation~(\ref{eq:byparts})
is suppressed by the time derivative of $f(q,s)$, whose size we estimate 
using the fitting formulae~(\ref{eq:fitting}) and (\ref{eq:f}) to be
\begin{equation}
\frac{\partial f(q,s)}{\partial s} \simeq  \frac{c \mu r^c}{(1+r^c)^2},
\end{equation}
and $\partial f/\partial s \geq 0$ for all $s$, meaning that the linear growth rate
cannot decrease in an Einstein--de Sitter cosmology.
For a given $q$ value, $\partial f/\partial s$ peaks at $r=1$, or 
$a=[1+y_{\rm FS}(q)] \ a_{\rm eq}$,
when the neutrinos transit from a non-clustering to a clustering dark matter,
thereby enhancing the (linear) gravitational source term.
Here, $\partial f/\partial s|_{r=1} \simeq c \mu/4 \simeq 0.1 f_\nu$.
Since the integrand is positive definite, we can conclude based on this estimate that
the absolute fractional error incurred by dropping the second term of the integral~(\ref{eq:byparts}) 
is bounded from above by $\sim 0.1 f_\nu$ (worst case:  $\kappa_a \simeq -1$, $c=1$, $d=2$, $q_1=q_2$);
the actual error is likely to be much less.
 Hence we shall proceed by keeping only the 
first term of equation~(\ref{eq:byparts}).

Of what remains of the time integral~(\ref{eq:byparts}), we retain as usual
only the non-decaying part, leading to
\begin{eqnarray}
\tilde{\psi}^{(2)}_a({\bm k},s) = \int d^3{\bm q}_1 d^3{\bm q}_2 
\frac{D_1(q_1,s) D_1(q_2,s)f(q_1,s) }{ \kappa_a+f(q_1,s)+f(q_2,s)} \nonumber \\
\hspace{35mm} \times \delta_D({\bm k}-{\bm q}_{12})  
 U^{-1}_{ab}  \gamma_{bcd} ({\bm q}_1,{\bm q}_2) \nonumber \\
\hspace{45mm}		Q_c^{(1)} ({\bm q}_1,s) Q_d^{(1)} ({\bm q}_2,s) \delta({\bm q}_1,s_0) \delta({\bm q}_2,s_0).
\end{eqnarray}
Rotating back to the original basis gives us
\begin{equation}
\psi^{(2)}_a({\bm k},s) \!= \! \int \! d^3{\bm q}_1 d^3{\bm q}_2 \delta_D({\bm k}-{\bm q}_{12}) 
Q^{(2)}_{a}({\bm q}_1,{\bm q}_2;s) \delta_1({\bm q}_1,s) \delta_1({\bm q}_2,s),
\end{equation}
where
\begin{equation}
Q^{(2)}_{a}({\bm q}_1,{\bm q}_2;s) \equiv \frac{U_{ab} U^{-1}_{bc}
 \gamma_{cde} ({\bm q}_1,{\bm q}_2) Q_d^{(1)} ({\bm q}_1,s) Q_e^{(1)} ({\bm q}_2,s)}
{\kappa_b+f(q_1,s)+f(q_2,s)}
\end{equation}
is the interaction kernel.

(iii) For $n=3$, the same procedure and approximation scheme yield
\begin{eqnarray}
\psi^{(3)}_a({\bm k},s) = \int  d^3{\bm q}_1 d^3{\bm q}_2 d^3{\bm q}_3 \delta_D({\bm k}-{\bm q}_{123}) 
\nonumber \\
 \hspace{30mm} \times 
 Q^{(3)}_{a}({\bm q}_1,{\bm q}_2,{\bm q}_3;s) \delta_1({\bm q}_1,s) \delta_1({\bm q}_2,s) \delta_1({\bm q}_3,s),
\end{eqnarray}
with 
\begin{eqnarray}
Q^{(3)}_{a}({\bm q}_1,{\bm q}_2,{\bm q_3};s) &\equiv& 
\frac{U_{ab} U^{-1}_{bc} }{\kappa_b+f(q_1,s)+ f(q_2,s)+f(q_3,s)}
\nonumber \\
&& 
\times \left[\gamma_{cde}({\bm q}_{12},{\bm q}_3) Q_d^{(2)}({\bm q}_1, {\bm q}_2;s)
Q_e^{(1)} ({\bm q}_3,s) \right.
 \nonumber \\
&& \hspace{7mm} +\left.\gamma_{cde}({\bm q}_1,{\bm q}_{23};s) Q_d^{(1)}({\bm q}_1,s)
Q_e^{(2)}({\bm q}_2,{\bm q}_3;s)\right].
\end{eqnarray}
The maximum absolute fractional error incurred by neglecting the time derivatives of $f(q,s)$
is again estimated at a negligible $\sim 0.1 f_\nu$ for the $n=3$ solution.

\section{General $n$th order solution and recursion relations\label{sec:nthorder}}

Generalising to the $n$th order, we obtain the approximate solution
\begin{eqnarray}
\label{eq:solution}
\psi^{(n)}_a({\bm k},\tau) = \int  d^3{\bm q}_1 \cdots d^3{\bm q}_n \delta_D({\bm k}-{\bm q}_{1\cdots n}) 
\nonumber \\
 \hspace{30mm} \times 
 Q^{(n)}_{a}({\bm q}_1,\cdots,{\bm q}_n;\tau) \delta_1({\bm q}_1,\tau) \cdots \delta_1({\bm q}_n,\tau),
\end{eqnarray}
where the interaction kernel $Q^{(n)}$ can be constructed from 
the recursion relation
\begin{eqnarray}
\label{eq:recursion}
Q^{(n)}_{a}({\bm q}_1,\cdots,{\bm q}_n;\tau) = \sigma^{(n)}_{ab}(q_1,\cdots,q_n;\tau)
\nonumber \\
 \hspace{40mm} \times 
 \sum_{m=1}^{n-1} \gamma_{bcd}({\bm q}_{1\cdots m},{\bm q}_{m+1 \cdots n}) 
Q_c^{(m)}({\bm q}_1,\cdots, {\bm q}_m;\tau) \nonumber \\
 \hspace{60mm} \times Q_d^{(n-m)} ({\bm q}_{m+1},\cdots, {\bm q}_n;\tau),
\end{eqnarray}
with
\begin{eqnarray}
\label{eq:sigma}
\sigma^{(n)}_{ab}(q_1,\cdots,q_n) &\equiv&
\frac{U_{ac}U^{-1}_{cb}}{\kappa_c+\omega^{(n)}(q_1,\cdots,q_n;\tau)}, \nonumber \\
&\doteq& \frac{1}{{\cal N}^{(n)}}\left[\begin{array}{cc}
		2\omega^{(n)}+1 & 2 \\
		3(1-f_\nu) & 2 \omega^{(n)} \end{array} \right],
\end{eqnarray}
and	
\begin{eqnarray}
\label{eq:sigmastuff}
 {\cal N}^{(n)}
(q_1,\cdots,q_n;\tau)  \equiv (2 \omega^{(n)}+3)(\omega^{(n)}-1)+3 f_\nu, \nonumber \\	
 \omega^{(n)}(q_1,\cdots,q_n;\tau)\equiv \sum_{m=1}^n f(q_m,\tau) 
= \sum_{m=1}^n \frac{\partial \ln D_1(q_m,\tau)}{\partial \ln a(\tau)}.
\end{eqnarray}
Equations~(\ref{eq:solution}) to~(\ref{eq:sigmastuff}) should be compared with, e.g.,
equations~(41) to (44), (83) and (84) of reference~\cite{Bernardeau:2001qr} 
for a CDM-only universe.

In the limit $f_\nu \to 0$, we have
$\omega^{(n)}(q_1,\cdots,q_n;\tau) \to n$,
so that the $\sigma^{(n)}$ matrix~(\ref{eq:sigma}) depends only on 
$n$ such as in equation~(84) of~\cite{Bernardeau:2001qr}.
This renders the 
interaction kernel~(\ref{eq:recursion}) into the standard expressions
$F_n({\bm q}_1,\cdots,{\bm q}_n)$ and $G_n({\bm q}_1,\cdots,{\bm q}_n)$ 
for a CDM-only universe given in equations~(43) and~(44) of \cite{Bernardeau:2001qr}, i.e.,
\begin{eqnarray}
Q^{(n)}_{1}({\bm q}_1,\cdots,{\bm q}_n;\tau) \to F_n({\bm q}_1,\cdots,{\bm q}_n), \nonumber \\
Q^{(n)}_{2}({\bm q}_1,\cdots,{\bm q}_n;\tau) \to G_n({\bm q}_1,\cdots,{\bm q}_n).
\end{eqnarray}
We caution at this point that the terms proportional to $f_\nu$ in equations~(\ref{eq:sigma}) and 
(\ref{eq:sigmastuff}) count only the explicit dependences on $f_\nu$;
the function $\omega^{(n)}$ also depends implicitly on 
$f_\nu$ through the linear growth function.

\subsection{Symmetrised kernels}

In practice we use the symmetrised kernels 
$\bar{Q}^{(n)}({\bm q}_1,\cdots,{\bm q}_n;\tau)$,
constructed by summing $Q^{(n)}({\bm q}_1,\cdots,{\bm q}_n;\tau)$
over all permutations of the momenta ${\bm q}_1,\cdots,{\bm q}_n$ and then
dividing by $n!$.
For future reference, we give here explicit expressions for 
$\bar{Q}^{(2)}({\bm q}_1,{\bm q}_2;\tau)$
and $\bar{Q}^{(3)}({\bm q}_1,{\bm q}_2.{\bm q}_3;\tau)$,
cast in a form as close to the standard CDM-only ones 
as possible:
\begin{eqnarray}
\label{eq:q2}
\bar{Q}^{(2)}_1({\bm q}_1,{\bm q}_2;\tau) = \frac{5}{7}  A_1  
+ \frac{1}{2} \frac{{\bm q}_1 \cdot {\bm q}_2}{q_1 q_2} \left[A_2 \frac{q_1}{q_2}
+ A_3 \frac{q_2}{q_1}\right] + \frac{2}{7} A_4
\frac{({\bm q}_1 \cdot {\bm q}_2)^2}{q_1^2 q_2^2}, \nonumber \\
\bar{Q}^{(2)}_2({\bm q}_1,{\bm q}_2;\tau) = \frac{3}{7}  C_1  
+ \frac{1}{2} \frac{{\bm q}_1 \cdot {\bm q}_2}{q_1 q_2} \left[ C_2 \frac{q_1}{q_2}
+  C_3 \frac{q_2}{q_1}\right] + \frac{4}{7} C_4
\frac{({\bm q}_1 \cdot {\bm q}_2)^2}{q_1^2 q_2^2}, 
\end{eqnarray}
with
\begin{eqnarray}
\label{eq:q2stuff}
A_1 =  \frac{7}{10}\sigma^{(2)}_{11}(q_1,q_2)[f(q_1)+f(q_2)],\nonumber \\
A_2 = f(q_2)[\sigma^{(2)}_{11}(q_1,q_2)+\sigma^{(2)}_{12}(q_1,q_2) f(q_1)], \nonumber \\
A_3 = f(q_1)[\sigma^{(2)}_{11}(q_1,q_2)+\sigma^{(2)}_{12}(q_1,q_2) f(q_2)], \nonumber \\
A_4 = \frac{7}{2}\sigma^{(2)}_{12} (q_1,q_2) f(q_1) f(q_2), \nonumber \\
C_1 =  \frac{7}{6}\sigma^{(2)}_{21}(q_1,q_2) [f(q_1)+f(q_2)],\nonumber \\
C_2 = f(q_2)[\sigma^{(2)}_{21}(q_1,q_2)+\sigma^{(2)}_{22}(q_1,q_2)f(q_1)],\nonumber \\
C_3 = f(q_1)[\sigma^{(2)}_{21}(q_1,q_2)+\sigma^{(2)}_{22}(q_1,q_2)f(q_2)],\nonumber \\
C_4 = \frac{7}{4}\sigma^{(2)}_{22} (q_1,q_2) f(q_1) f(q_2),
\end{eqnarray}
where we have dropped the $\tau$ label in $\sigma^{(2)}(q_1,q_2;\tau)$ and
$f(q,\tau)$
for convenience.
The factors  $A_{1,2,3,4}$ and $C_{1,2,3,4}$ have been defined so that they tend to unity 
as $f_\nu \to 0$.

The symmetrised form of $Q^{(3)}$ can be written as
\begin{eqnarray}
\label{eq:q3}
\bar{Q}^{(3)}_1({\bm q}_1,{\bm q}_2,{\bm q}_3;\tau) = \frac{1}{3} \left\{
\frac{7}{18} X_{123} \left(\frac{{\bm q}_{123} \cdot {\bm q}_3}{q_3^2} \right) \bar{Q}^{(2)}_1({\bm q}_1,{\bm q}_2;\tau)
\right. \nonumber \\
\hspace{10mm} + \left[\frac{7}{18} Y_{123} \left(\frac{{\bm q}_{123} \cdot {\bm q}_{12}}{q_{12}^2} \right)
+ \frac{1}{9} Z_{123} \left(\frac{q_{123}^2 ({\bm q}_{12} \cdot {\bm q}_{3})}{q_{12}^2 q_3^2} \right) \right]
\bar{Q}^{(2)}_2({\bm q}_1,{\bm q}_2;\tau) \nonumber \\
 \hspace{10mm} \left.+ 
\ {\rm cyclic \ permutations} \vphantom{\frac{1}{3}} \right\},
\end{eqnarray}
where
\begin{eqnarray}
\label{eq:xyz}
 X_{123} \equiv X(q_1,q_2,q_3) = \frac{18}{7} \sigma^{(3)}_{11}(q_1,q_2,q_3)f(q_3), \nonumber \\
Y_{123} \equiv Y(q_1,q_2,q_3)  =\frac{18}{7} \sigma^{(3)}_{11}(q_1,q_2,q_3), \nonumber \\
 Z_{123} \equiv Z(q_1,q_2,q_3) =9 \sigma^{(3)}_{12}(q_1,q_2,q_3)f(q_3).
\end{eqnarray}
Again, $X_{123},Y_{123},Z_{123} \to 1$ as $f_\nu\to 0$.
To obtain $\bar{Q}_2^{(3)}({\bm q}_1,{\bm q}_2,{\bm q}_3;\tau)$, simply replace 
$\sigma^{(3)}_{11}$ with $\sigma^{(3)}_{21}$, and $\sigma^{(3)}_{12}$ with $\sigma^{(3)}_{22}$
in the expressions for $X_{123}$, $Y_{123}$, and $Z_{123}$.

\section{Power spectra\label{sec:spectra}}

We are interested in the total matter power spectrum $P(k,\tau)$.  As in the 
case for the linear power spectrum~(\ref{eq:linpow}), it can be expressed as
a weighted sum of the CDM+baryon and the neutrino density contrast
auto- and cross-correlation power spectra,
\begin{equation}
P(k,\tau) = f^2_{cb} P_{cb}(k,\tau) + 2 f_{cb} f_\nu P_{cb \nu} (k,\tau) + f_\nu^2 P_{\nu}(k, \tau),
\end{equation}
where 
\begin{eqnarray}
 P_{cb}(k,\tau) \delta_D({\bm k}+{\bm k}') \equiv \langle \delta({\bm k},\tau) \delta({\bm k}',\tau) \rangle, \nonumber \\
 P_{\nu}(k,\tau) \delta_D({\bm k}+{\bm k}') \equiv \langle \delta^\nu({\bm k},\tau) \delta^\nu({\bm k}',\tau) 
\rangle, \nonumber \\
 P_{cb \nu}(k,\tau) \delta_D({\bm k}+{\bm k}') \equiv \langle \delta({\bm k},\tau) \delta^{\nu}({\bm k}',\tau) 
\rangle.
\end{eqnarray}
Assuming Gaussian initial conditions, these various power spectra, up 
to one-loop corrections, are given by~\cite{bib:vishniac}
\begin{eqnarray}
 P_{cb}(k,\tau) = P^L_{cb}(k,\tau) + [P^{(22)}_{cb}(k,\tau)+2 P^{(13)}_{cb}(k,\tau)], \nonumber \\
 P_{\nu} (k,\tau) = P^L_{\nu}(k,\tau), \nonumber \\
 P_{c b\nu}(k,\tau) = P^L_{cb \nu}(k,\tau) + P^{(13)}_{cb\nu}(k,\tau), 
\end{eqnarray}
taking the neutrino density perturbations to be linear at all times.

The one-loop terms for the CDM+baryon auto-correlation are
\numparts
\begin{eqnarray}
P^{(22)}_{cb}(k,\tau) &=& 2 \int d^3 {\bm q} \ [\bar{Q}_1^{(2)}({\bm k}-{\bm q},{\bm q};\tau)]^2 \ 
P^L_{cb}(|{\bm k}-{\bm q}|,\tau)  P^L_{cb}(q,\tau),\label{eq:22} \\
P^{(13)}_{cb}(k,\tau) &=& 3 \ P^L_{cb}(k,\tau)  \int d^3 {\bm q} \ \bar{Q}_1^{(3)}({\bm k},{\bm q},-{\bm q};\tau) 
  P^L_{cb}(q,\tau), \label{eq:13}
\end{eqnarray}
\endnumparts
where the prefactors ``2'' and ``3'' arise from summing all possible groupings of 
$\langle \delta^I({\bm k}) \delta^I({\bm k}') \rangle$ pairs to give
$\langle \delta^I({\bm k}) \delta^I({\bm k}') \delta^I({\bm k}'') \delta^I({\bm k}''')\rangle$
(e.g., \cite{Fry:1993bj}).
The cross-correlation  between CDM+baryons and neutrinos also contains a one-loop correction,
\begin{equation}
\label{eq:nu13}
P^{(13)}_{cb\nu}(k,\tau) =   3 \ P^L_{c b\nu}(k,\tau)  \int d^3 {\bm q} 
\ \bar{Q}_1^{(3)}({\bm k},{\bm q},-{\bm q};\tau)  P^L_{cb}(q,\tau),
\end{equation}
which is identical to the $P^{(13)}_{cb}(k,\tau)$ correction to the CDM+baryon auto-correlation except for the prefactor
$P^L_{cb \nu}(k,\tau)$ instead of $P^L_{c b}(k,\tau)$.  This term is missing from Saito~{\it et~al.}'s
formulation, which assumed explicitly $P_{c b\nu}(k,\tau)= P^L_{cb\nu}(k,\tau)$~\cite{Saito:2008bp}.  
This assumption is not self-consistent, since a one-loop correction must be present 
in the CDM+baryon--neutrino cross-correlation, if the CDM+baryon density contrast has indeed been calculated to
third order in perturbative expansion.

\subsection{Explicit forms\label{sec:explicit}}

Using equations~(\ref{eq:q2}) and (\ref{eq:q2stuff}) to evaluate
$\bar{Q}_1^{(2)}({\bm k}-{\bm q},{\bm q};\tau)$, and defining $x \equiv {\bm k} \cdot {\bm q}/(kq)$,
$r \equiv q/k$, and  $\eta \equiv \sqrt{1+r^2 - 2 r x}$, we find for the ``22'' correction term~(\ref{eq:22})
\begin{eqnarray}
\label{eq:22explicit}
P_{cb}^{(22)} (k,\tau) =\frac{2 \pi k^3}{98} \int_0^{\infty} d r P_{cb}^L (kr,\tau)
\int_{-1}^1 d x P_{cb}^L(k \sqrt{1+r^2 - 2 r x},\tau) \nonumber \\
 \hspace{50mm} \times \frac{[3 S_1 r+ 7 (S_2+S_3 r^2) x - 10 S_4 r x^2]^2}{(1+r^2 - 2 rx)^2},
\end{eqnarray}
with
\begin{eqnarray}
\label{eq:s}
S_1=
\frac{7}{3}  f(k \eta) [\sigma^{(2)}_{11}(k \eta,kr) 
- \sigma^{(2)}_{12} (k \eta,kr) f(kr) ], \nonumber \\
S_2 = f(kr)[\sigma^{(2)}_{11}(k \eta,kr)+\sigma^{(2)}_{12}(k \eta,kr) f(k \eta)],
\nonumber \\
S_3 = \sigma^{(2)}_{11}(k\eta,kr) [f(kr)-f(k \eta)], \nonumber \\
S_4  = 
\frac{7}{5}  \sigma^{(2)}_{11}(k\eta,kr) f(kr),
\end{eqnarray}
and the functions $f(q)$ and $\sigma^{(n)}_{ab}(q_1,\cdots,q_n)$ are defined 
in equations~(\ref{eq:ffunc}) and (\ref{eq:sigma}) respectively.
In the $f_\nu\to0$ limit, we have $S_{1,2,4}\to 1$ and $S_3\to0$, which is the standard CDM-only result first given
in references~\cite{Suto:1990wf,Makino:1991rp} and adopted in the analysis of~\cite{Saito:2008bp}.

Similarly, evaluating $\bar{Q}_1^{(3)}({\bm k},{\bm q},-{\bm q};\tau)$ with the 
aid of equations~(\ref{eq:q3}) and (\ref{eq:xyz}), we obtain
for the ``13'' terms~(\ref{eq:13}) and (\ref{eq:nu13})
\begin{equation}
\label{eq:13explicit}
2 P_{cb,cb\nu}^{(13)} (k,\tau) 
= \frac{2 \pi k^3}{63} P_{cb,cb\nu}^L (k,\tau)  \int_0^{\infty}  \! dr 
P_{cb}^L (kr,\tau)  \int_{-1}^{1}  \! dx
 \left[\frac{V}{\eta^2} +  W \right],
\end{equation}
where
\begin{eqnarray}
V = \left[6 I_1 r - 7 (I_2  \!+\!I_3 r^2) x + 8 I_4 r x^2 \right] \!
\left[(7M\!-\!2R) r - (7 M r^2 \!-\! 2 R) x \right],\nonumber \\
 W = 7 L \left[10 H_1 r x - 7 (H_2 + H_3 r^2) x^2 + 4 H_4 r x^3\right],
\end{eqnarray}
and
\begin{eqnarray}
H_1=  \frac{7}{10}\sigma^{(2)}_{11}(k,kr)[f(k)+f(kr)],\nonumber \\
H_2 =  f(kr)[\sigma^{(2)}_{11}(k,kr)+\sigma^{(2)}_{12}(k,kr) f(k)], \nonumber \\
H_3 = f(k)[\sigma^{(2)}_{11}(k,kr)+\sigma^{(2)}_{12}(k,kr) f(kr)], \nonumber \\
H_4 = \frac{7}{2}\sigma^{(2)}_{12} (k,kr) f(k) f(kr), \nonumber \\
I_1 = \frac{7}{6}\sigma^{(2)}_{21}(k,kr) [f(k)+f(kr)],\nonumber \\
I_2 = f(kr)[\sigma^{(2)}_{21}(k,kr)+\sigma^{(2)}_{22}(k,kr)f(k)],\nonumber \\
I_3=f(k)[\sigma^{(2)}_{21}(k,kr)+\sigma^{(2)}_{22}(k,kr)f(kr)],\nonumber \\
I_4 =  \frac{7}{4}\sigma^{(2)}_{22} (k,kr) f(k) f(kr),\nonumber \\
L= \frac{18}{7} \sigma^{(3)}_{11}(k,kr,kr) f(kr), \nonumber \\
M= \frac{18}{7} \sigma^{(3)}_{11}(k,kr,kr), \nonumber \\
 R = 9 \sigma^{(3)}_{12}(k,kr,kr) f(kr).
\end{eqnarray}
Again, these factors have been defined so that $H_{1,2,3,4}$, $I_{1,2,3,4}$, and $L,M,R$
tend to unity in the $f_\nu = 0$ limit.  
Since these factors have no $x$-dependence, we can
further simplify equation~(\ref{eq:13explicit}) by performing the integration over $x$ to obtain
\begin{eqnarray}
\label{eq:xintegral}
\int_{-1}^{1}  dx \left[\frac{V}{\eta^2} +  W \right]
&=&\frac{1}{4 } \left[ \frac{12 T_1}{r^2}-158T_2+100 T_3 r^2 - 42 T_4 r^4  \right. \nonumber \\
&& \hspace{-10mm} \left. + \frac{3}{r^3} (r^2-1)^2 (T_5 r^2 - T_6) (7 T_7 r^2+2 T_8)
\ln \left|\frac{1+r}{1-r}\right|
\right],
\end{eqnarray}
where
\begin{eqnarray}
\label{eq:t}
T_1 &=& 
21 \sigma_{12}^{(3)}(k,kr,kr) \sigma_{21}^{(2)}(k,kr) f^2(kr), \nonumber \\
T_2 &=& 
\frac{42}{79} f(kr) \left\{\sigma_{11}^{(3)}(k,kr,kr)\left[4 \sigma_{11}^{(2)}(k,kr) f(kr) \right.  \right. \nonumber \\
&& \hspace{45mm} \left. + 
4 \sigma_{12}^{(2)}(k,kr) f(k) f(kr) - 3 \sigma_{21}^{(2)}(k,kr)\right]  \nonumber \\
&& \hspace{20mm} + \sigma_{12}^{(3)}(k,kr,kr)  \left[ \sigma_{21}^{(2)}(k,kr)[3 f(k)+9 f(kr)]  \right. \nonumber \\
&& \hspace{60mm} \left. \left.+
4 \sigma_{22}^{(2)}(k,kr) f(k) f(kr) \right] \right\},
\nonumber \\
T_3&=&  \frac{21}{25} \left\{ -3 \sigma_{12}^{(3)}(k,kr,kr) \sigma_{21}^{(2)}(k,kr) f(k)  f(kr)\right. \nonumber \\
&& \hspace{10mm} +\sigma_{11}^{(3)}(k,kr,kr) \left[ \sigma_{21}^{(2)}(k,kr) [3 f(kr)+9 f(k)] \right. \nonumber \\
&& \hspace{45mm} +4 [\sigma_{22}^{(2)}(k,kr)-\sigma_{11}^{(2)}(k,kr)] f(k)  f(kr) 
\nonumber \\
&& \hspace{45mm} \left. \left. -4 \sigma_{12}^{(2)}(k,kr) f(k)f^2(kr)   \right] \right\}, \nonumber \\
T_4 &=& 
6 \sigma_{11}^{(3)}(k,kr,kr) \sigma_{21}^{(2)}(k,kr) f(k),\nonumber \\
T_5 &=& f(k), \nonumber \\
T_6 &=& f(kr), \nonumber \\
T_7 &=& 6 \sigma^{(3)}_{11}(k,kr,kr) \sigma^{(2)}_{21}(k,kr), \nonumber \\
T_8 &=& 21 \sigma^{(3)}_{12}(k,kr,kr) \sigma^{(2)}_{21}(k,kr) f(kr),
\end{eqnarray}
and $T_{1,\cdots,8} \to 1$ as $f_\nu \to 0$.  The $f_\nu =0$ limit of equations~(\ref{eq:13explicit}) and
(\ref{eq:xintegral}) was first derived in~\cite{Suto:1990wf,Makino:1991rp} and then used in~\cite{Saito:2008bp}.

\section{Results and discussions\label{sec:discuss}}

\subsection{One-loop corrected power spectra}

\begin{figure}[t]
\includegraphics[width=15.5cm]{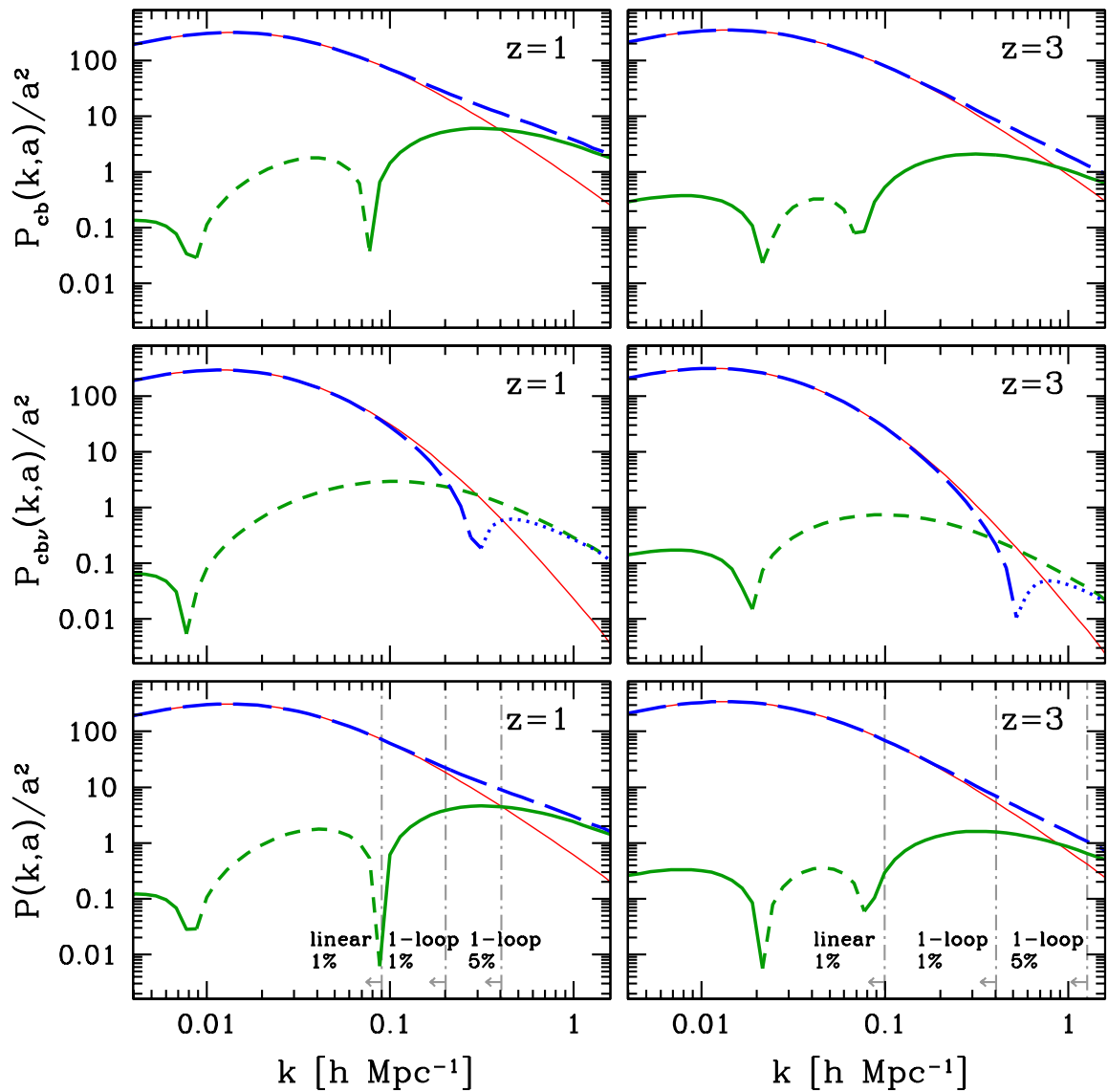}
\caption{Contributions to the total matter power spectrum (units: $h^3 \ {\rm Mpc}^{-3}$) at 
$z=1$ (left) and $z=3$ (right) for a $\Lambda$CHDM cosmology with
$f_\nu=0.1$.  {\it Top}: CDM+baryon auto-correlation $P_{cb}(k,\tau)$.   
{\it Middle}: CDM+baryon--neutrino cross-correlation $P_{cb\nu}(k,\tau)$. 
{\it Bottom}: Total matter power spectrum $P(k,\tau)$.
In all cases the linear contribution is shown in red/thin solid,
the one-loop correction in green/thick solid (green/short dash when negative),
and their sum in blue/long dash (blue/dotted when negative).
All spectra have been divided by $a^2$ to facilitate comparison.
The three vertical lines demarcate, from left to right, 
the upper limits in $k$ for which the linear matter power spectrum is accurate to
1\% or less, the one-loop corrected spectrum to 1\% or less, and the
one-loop corrected spectrum to 5\% or less.
\label{fig:explicit}}
\end{figure}

Figure~\ref{fig:explicit} shows various contributions to the total matter power spectrum 
at $z=1$ and $z=3$ in a $\Lambda$ mixed cold+hot dark matter ($\Lambda$CHDM) cosmology,
assuming a total  matter density $\Omega_m h^2 = 0.14$, neutrino fraction 
$f_\nu=0.1$, Hubble rate $h=0.72$, and scalar spectral index $n_s=0.963$.  The total matter power 
spectrum has been normalised at the pivot scale  $k_0=0.002 \ {\rm Mpc}^{-1}$ 
to the vanilla best-fit  from the 
Wilkinson Microwave Anisotropy Probe five-year data,
i.e., the best-fit amplitude of the curvature perturbations 
$\Delta_{\cal R}^2=2.4 \times 10^{-9}$~\cite{Dunkley:2008ie}.
Note that at the chosen redshifts the universe is very nearly described by an Einstein--de Sitter model.  However, 
the vacuum energy $\Lambda$ term still has some residual effects, particularly on the logarithmic 
derivative of the linear growth function, $f(k,\tau)$.
It is therefore necessary to rescale $f(k,\tau)$ so that $f(k,\tau) \to 1$  as $k\to 0$, in order for
the expressions~(\ref{eq:22explicit}) to (\ref{eq:13explicit}),
(\ref{eq:xintegral}) and (\ref{eq:t}) to be applicable.

Observe that the $P^{(22)}_{cb}(k,\tau)$ term derives from a positive definite integrand 
and is therefore always positive. The $P^{(13)}_{cb}(k,\tau)$ term, on the other hand, is negative at most $k$ values.  
The net effect of their sum on
the CDM+baryon auto-correlation is that the power spectrum is  first suppressed at 
$k \sim 0.01 \to 0.1 \ h \ {\rm Mpc}^{-1}$, and then enhanced as 
we move beyond $k \sim 0.1 \ h \ {\rm Mpc}^{-1}$.

Contrastingly, the cross-correlation spectrum between the CDM+baryon and neutrino components 
receives a one-loop correction only from $P^{(13)}_{cb\nu}(k,\tau)$.
Like the $P^{(13)}_{cb}(k,\tau)$ term, $P^{(13)}_{cb\nu}(k,\tau)$ is mostly negative.
This causes the already free-streaming-suppressed  power spectrum---recall 
the neutrino linear growth function falls off like $k^{-2}$ at $k \gg k_{\rm FS}$, see 
equation (\ref{eq:nufitting})---to be further suppressed, before turning  negative at $k \sim 0.2 \ h \ {\rm Mpc}^{-1}$.  
Beyond $k \sim 0.2 \ h \ {\rm Mpc}^{-1}$, however, the one-loop correction dominates over the linear term, resulting in an 
enhancement in the magnitude of $P_{cb\nu}(k,\tau)$, but in the negative direction.

The net correction to the total matter power spectrum follows essentially the same trend as 
the correction to $P_{cb}(k,\tau)$, beginning with a small suppression at $k \lwig 0.1 \ h \ {\rm Mpc}^{-1}$, 
and culminating in an enhancement at $k \gwig 0.1 \ h \ {\rm Mpc}^{-1}$.

\subsection{Regions of validity}

Perturbation theory is not expected to describe reality at very large $k$ values, 
since any perturbative expansion must break down when the evolution of structures becomes fully nonlinear.  
A good rule of thumb is to take 
as the valid regime
the range of $k$ values at which the one-loop correction is smaller than the 
linear contribution~\cite{Jain:1993jh}.
More recent analyses show that for a $\Lambda$CDM cosmology, the one-loop corrected 
power spectrum deviates by less than 1\% from  $N$-body simulation results
provided that the dimensionless power spectrum, defined in this work as
$\Delta^2(k,\tau) \equiv 4 \pi k^3 P(k,\tau)$, does not exceed $\sim 0.4$~\cite{Jeong:2006xd}.  
Applying this criterion to
the total matter power spectrum in our $\Lambda$CHDM scenario, we expect our one-loop 
corrected power spectrum to be accurate to better than 1\%
for $k \lwig 0.2 \ h \ {\rm Mpc}^{-1}$ at $z=1$ and $k \lwig 0.4 \ h \ {\rm Mpc}^{-1}$ at $z=3$.
For completeness we also estimate a 5\% accurate region following
figure~2 of~\cite{Jeong:2006xd}: $k \lwig 0.4 \ h \ {\rm Mpc}^{-1}$ at $z=1$, and 
$k \lwig 1 \ h \ {\rm Mpc}^{-1}$ at $z=3$.

It is also interesting to ask up to what values of $k$ is {\it linear} theory expected to provide an 
accurate description of the matter power spectrum.
Comparing the total matter power spectrum computed from linear theory to that including the one-loop correction, 
we find 1\% agreement between the two only at $k \lwig 0.09 \ h \ {\rm Mpc}^{-1}$ for $z=1$ and 
$k \lwig 0.1 \ h \ {\rm Mpc}^{-1}$ for $z=3$.
From this we conclude that the one-loop correction improves on linear theory to better than 1\% accuracy 
in the $k$ ranges
\begin{eqnarray}
0.09 \lwig k/(h \ {\rm Mpc}^{-1}) \lwig 0.2, & \qquad & z = 1,\nonumber \\
0.1 \lwig k/(h \ {\rm Mpc}^{-1}) \lwig 0.4, & \qquad &  z = 3.
\end{eqnarray}
Figure~\ref{fig:explicit} shows the various regions of validity discussed in this section.

\subsection{Suppression due to neutrino free-streaming}

\begin{figure}[t]
\includegraphics[width=15.5cm]{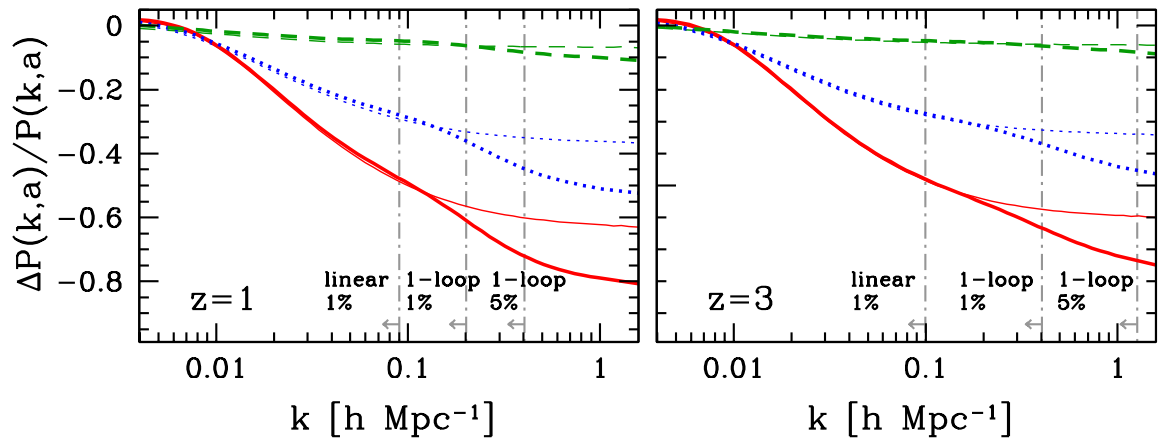}
\caption{Relative differences between the total matter power spectra
for a pure $\Lambda$CDM cosmology and  three $\Lambda$CHDM models
with $f_\nu=0.1$ (red/solid), 0.05 (blue/dotted), and 0.01 (green/dash)
at $z=1$ (left) and $z=3$ (right).  Thick lines indicate results 
including the one-loop correction, while the linear results 
are represented by the thin lines.  The three vertical lines indicate 
the maximum $k$ values at which the linear and the one-loop corrected matter
power spectra are accurate to better than 1\% and 5\%.\label{fig:suppression}}
\end{figure}

Figure~\ref{fig:suppression} shows the suppression in the total
matter power spectrum due to neutrino hot dark matter for three 
$\Lambda$CHDM models with $f_\nu=0.1,0.05,0.01$, relative to the case with
$f_\nu=0$, i.e., 
\begin{equation}
\frac{\Delta P(k,a)}{P(k,a)} \equiv \frac{P_{f_\nu \neq0}(k,a)-P_{f_\nu=0}(k,a)}{P_{f_\nu=0}(k,a)}.
\end{equation}
Observe that the relative decrease of small scale power due to neutrino free-streaming
in general exceeds the amount of suppression predicted by linear theory,
once nonlinear corrections have been included.
Within the 5\% accurate region, it is clear that the suppression easily exceeds 
the canonical linear suppression factor of $\sim 8 f_\nu$.%
\footnote{The linear suppression factor reaches a maximum of $\sim 8 f_\nu$ 
only for small values of $f_\nu$.  For, e.g., $f_\nu=0.2$, the linear 
suppression factor is $\sim 4.5 f_\nu$~\cite{Kiakotou:2007pz}.}
This enhanced suppression has been observed in multi-component 
$N$-body simulations~\cite{Brandbyge:2008rv}, which support
an asymptotic suppression factor of $\sim 9.8 f_\nu$.
Furthermore, compared to the linear results, 
the one-loop correction
leads to a  small increase in relative power at just below 
$k \sim 0.1 \ h \ {\rm Mpc}^{-1}$, before the enhanced suppression sets in,
 a feature that seems to be present also in 
figure~4 of~\cite{Brandbyge:2008rv}

A more rigorous comparison between our perturbation theory results
and the $N$-body  results of~\cite{Brandbyge:2008rv} is 
not possible at present, since our results are valid only for $z\gwig 1$, 
while reference~\cite{Brandbyge:2008rv} gives their results at $z=0$.

\begin{figure}[t]
\includegraphics[width=15.5cm]{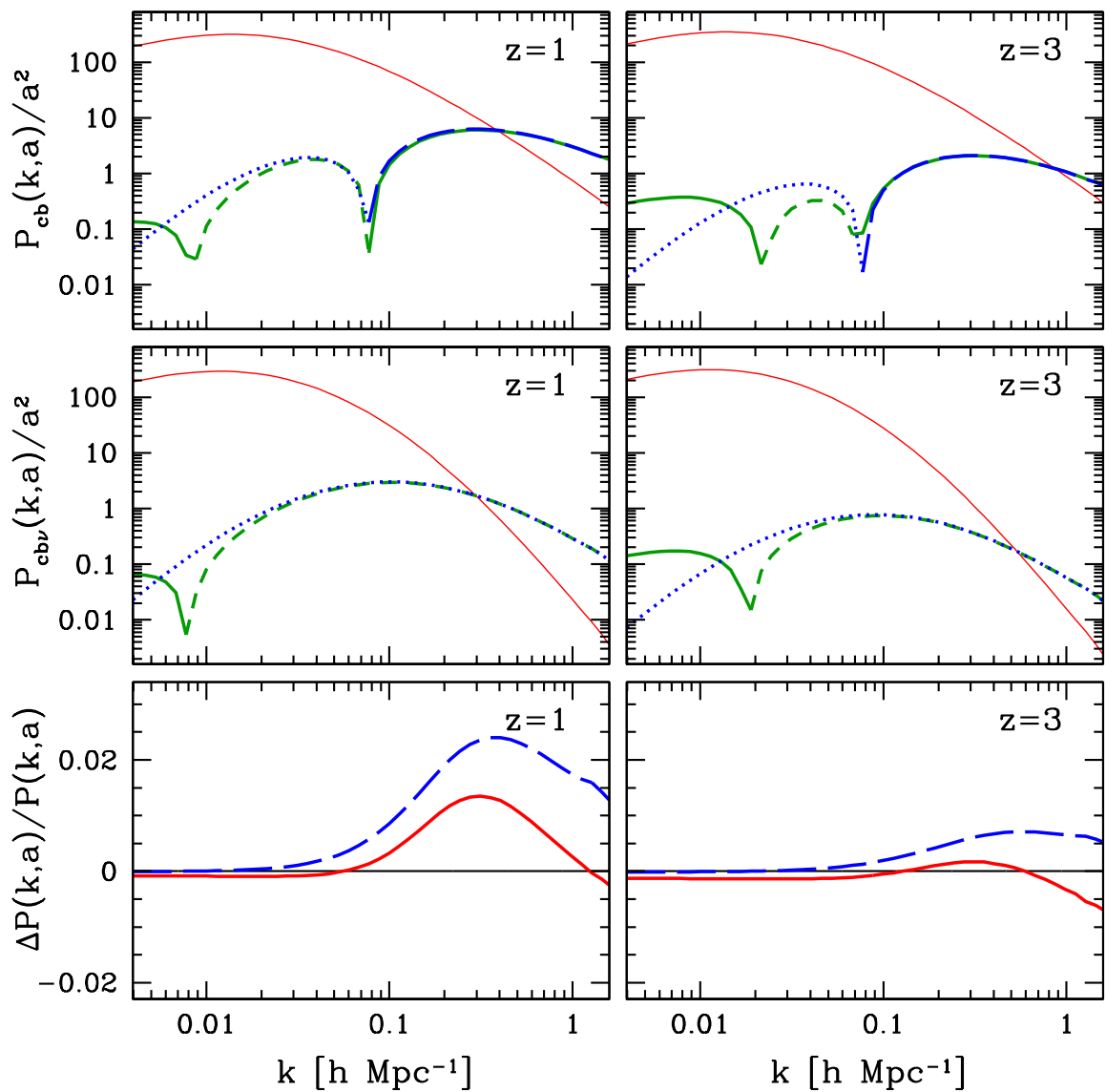}
\caption{One-loop corrections to the CDM+baryon auto-correlation  (top)
and the CDM+baryon--neutrino cross-correlation  (middle) 
at $z=1$ (left) and $z=3$ (right) for a $\Lambda$CHDM cosmology with
$f_\nu=0.1$.  In the top and middle rows, the red/thin solid lines
indicate the linear contribution, while
the green/thick solid lines denote the one-loop correction
(green/short dash when negative).  
The blue/long dash lines represent an approximation to
the one-loop correction (blue/dotted when negative), computed
by setting $S_{1,2,4}=T_{1,\cdots,8}=1$ and $S_3=0$.
The fractional error incurred in the total
matter power spectrum by this approximation is shown in the bottom row in 
red/solid.  Also plotted
in the bottom row in blue/long dash
is the fractional error incurred by neglecting the one-loop 
correction to the CDM+baryon--neutrino cross-correlation.
\label{fig:compare}}
\end{figure}

\subsection{Further approximations?}

There are two sources of deviation from the standard CDM-only case in the one-loop correction
terms~(\ref{eq:22explicit}), (\ref{eq:13explicit}) and (\ref{eq:xintegral}).
The first is encapsulated in the linear power spectra $P_{cb}^L(k,\tau)$ and $P_{c b \nu}^L(k,\tau)$;
the second in the factors $S_{1,\cdots,4}$ and $T_{1,\cdots,8}$ defined in
equations~(\ref{eq:s}) and (\ref{eq:t}), which depend on the linear growth functions.  
The linear power spectra are readily 
calculable with a sophisticated Boltzmann code such as {\tt CAMB}.
With some extra but minor effort the same is true also for the factors 
$S_{1,\cdots,4}$ and $T_{1,\cdots,8}$.
Nonetheless, it is tempting to simply assume $S_{1,2,4}=T_{1,\cdots,8}=1$ and $S_3=0$, 
as was done in~\cite{Saito:2008bp}.
In this section we examine the 
validity of this approximation.

\begin{figure}[t]
\includegraphics[width=15.5cm]{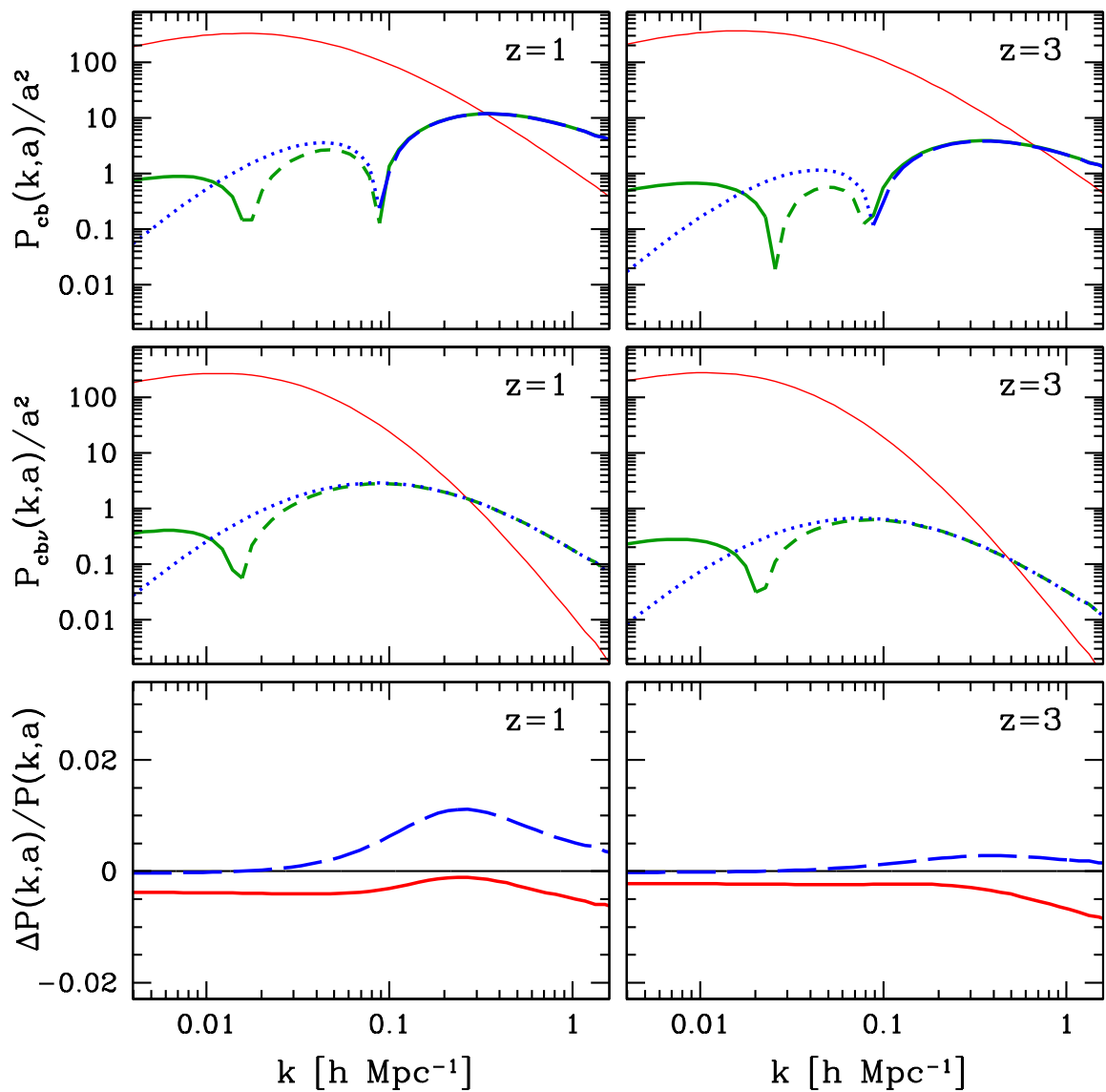}
\caption{Same as figure~\ref{fig:compare}, but for $f_\nu=0.05$.
\label{fig:compare1}}
\end{figure}

\begin{figure}[t]
\includegraphics[width=15.5cm]{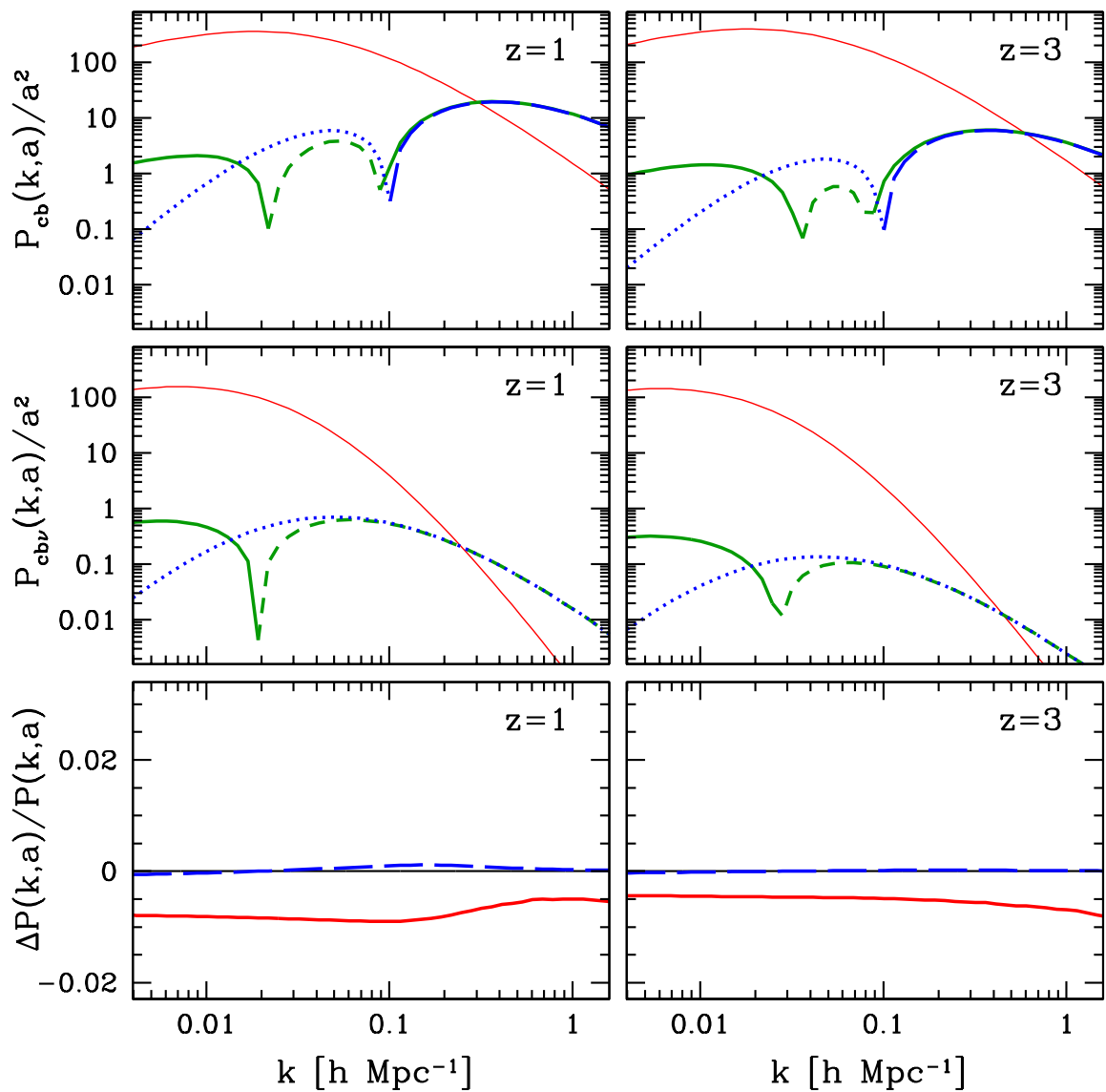}
\caption{Same as figure~\ref{fig:compare}, but for $f_\nu=0.01$.
\label{fig:compare2}}
\end{figure}

Figures~\ref{fig:compare}, \ref{fig:compare1} and \ref{fig:compare2}
show the correction terms
$P_{cb}^{(22)}+2 P_{cb}^{(13)}$ and $P_{cb\nu}^{(13)}$
computed under the assumption of
$S_{1,2,4}=T_{1,\cdots,8}=1$ and $S_3=0$ for three $\Lambda$CHDM 
models with $f_\nu=0.1,0.05,0.01$ respectively.  Compared with 
their correct forms, we find deviations as large as a
factor of twenty, sometimes even accompanied by a sign flip.  However,
such large deviations are confined to a region at
$k \ll 0.1 \ h \ {\rm Mpc}^{-1}$,
where the one-loop correction terms are in any case subdominant 
to the linear contribution; the net fractional error incurred in 
the total matter power spectrum turns out to never exceed the 1\% level
for our choices of $f_\nu$ and redshifts. Future cosmological probes will 
generally require an accuracy of $\sim 1$\% in the 
matter power spectrum in order not to bias parameter estimation. 
On this basis, we conclude that setting $S_{1,2,4}=T_{1,\cdots,8}=1$ 
and $S_3=0$ is a tolerable simplification.

Also shown in figures~\ref{fig:compare}, \ref{fig:compare1} and \ref{fig:compare2}
are the consequences of dropping the $P_{cb\nu}^{(13)}(k,\tau)$ correction 
to the CDM+baryon--neutrino cross-correlation.
As expected, the importance of this term increases as we increase $f_\nu$. 
For $f_\nu=0.1$, the maximum contribution of $P_{cb\nu}^{(13)}(k,\tau)$
to the $z=1$ total matter power spectrum is 2.5\% at $k \sim 0.4 \ h \ {\rm Mpc}^{-1}$.  
For $f_\nu=0.05$, the contribution is essentially halved.  Thus 
for $f_\nu <0.05$, dropping the  $P_{cb\nu}^{(13)}(k,\tau)$
 term will unlikely bias our results.
Nonetheless, $P_{cb\nu}^{(13)}(k,\tau)$ can be computed at essentially no extra expense to the user
because of its similarity to the (mandatory) $P_{cb}^{(13)}(k,\tau)$ contribution, cf.\
equation~(\ref{eq:13explicit}).  Hence
there is no real need to resort to approximations in this instance.

\section{Conclusions\label{sec:conclusions}}

In this paper we have presented the first rigorous and systematic derivation of the 
one-loop correction to the large scale matter power spectrum in a mixed dark matter
cosmology with subdominant massive neutrino hot dark matter.

Beginning with the relevant equations of motion, we find that by invoking an ``adiabatic'' 
approximation, accurate to better than 
$\sim 0.1 f_\nu$, higher order corrections to the CDM+baryon density contrast and velocity field
can be rendered into a form nearly identical to that for a pure-CDM cosmology.
The interaction kernels and their recursion relations also exhibit striking similarities
to their standard CDM-only counterparts, but contain additional dependences on the 
neutrino energy density fraction $f_\nu$ and the linear growth functions of the 
incoming wavevectors.  These results, generalised to $n$th order in perturbative expansion,
are summarised in equations~(\ref{eq:solution}) to (\ref{eq:sigmastuff}).

Using these approximate solutions we compute the usual ``22'' and ``13'' 
one-loop correction terms to the matter power spectrum.
As in the standard CDM-only case, these correction
terms take the form of integrals over the wavevector ${\bm q}$ of the linear 
power spectrum $P^L(q,\tau)$ multiplied by the interaction kernels. 
In addition to the corrections to the CDM+baryon auto-correlation,
we also find a one-loop correction term for the cross-correlation 
between the CDM+baryon and the neutrino components which was previously neglected.
These correction terms
appear in their evaluated and most simplified form in 
equations~(\ref{eq:22explicit}) to (\ref{eq:13explicit}),
(\ref{eq:xintegral}) and (\ref{eq:t}).

Evaluating these expressions numerically, we find that nonlinear corrections to the large
scale matter power spectrum can enhance the suppression  of 
small scale power due to neutrino free-streaming relative to the 
$f_\nu=0$ case to beyond the 
canonical linear suppression factor of $\sim 8 f_\nu$.
This enhanced suppression has been observed in multi-component
$N$-body simulations~\cite{Brandbyge:2008rv}.

As said, the interaction kernels contain hitherto unaccounted 
dependences on $f_\nu$ and the linear growth functions.  Neglecting these dependences in principle generates 
large deviations in the one-loop corrections.  However,
since these deviations occur at wavenumbers 
at which the linear contribution dominates over the correction terms, their 
 net effect on the total matter power spectrum never exceeds 1\%.
Future cosmological probes will require an accuracy of $\sim 1$\%
in the matter power spectrum in order not to bias parameter estimation. 
We have thus verified the validity of
the approach of~\cite{Saito:2008bp}.

An important assumption in our present treatment is that the neutrino density 
perturbations have been taken to remain linear at all times.  Although 
for realistic values of $f_\nu$ this can be justified by $N$-body simulation 
results~\cite{Brandbyge:2008rv},
a truly complete analysis of higher order corrections to the clustering 
statistics of the large scale structure distribution in the presence of massive
neutrinos should include also a proper account of nonlinear neutrino evolution.
We defer this investigation to a future publication.

Finally, as noted in reference~\cite{Saito:2008bp}, although higher order 
perturbation theory appears at first glance to have a limited range of validity---we expect 
our one-loop corrections to improve on linear theory to better than 1\% accuracy
in the region $0.1 \lwig k/(h \ {\rm Mpc}^{-1}) \lwig 0.4$ at $z=3$, it does enable an approximate 
factor of four increase in the maximum usable wavenumber in a data set.
This is equivalent to a factor of 64 gain in the number of independent Fourier modes, or
an eight-fold gain in statistical power for a fixed survey volume.
Such an improvement is no small feat, and may very well be just what we need  
to detect neutrino dark matter.

\ack

${\rm Y}^3$W thanks Jacob Brandbyge, Steen Hannestad and Georg Raffelt 
for useful discussions and/or comments on the manuscript.


\section*{References}

\begin{thebibliography}{99}


\bibitem{Amsler:2008zz}
  C.~Amsler {\it et al.}  [Particle Data Group],
  ``Review of particle physics,''
  Phys.\ Lett.\  B {\bf 667} (2008) 1.

\bibitem{Bond:1980ha}
  J.~R.~Bond, G.~Efstathiou and J.~Silk,
  ``Massive neutrinos and the large-scale structure of the universe,''
  Phys.\ Rev.\ Lett.\  {\bf 45} (1980) 1980.


\bibitem{Doroshkevich:1980zs}
  A.~G.~Doroshkevich, Y.~B.~Zeldovich, R.~A.~Sunyaev and M.~Khlopov,
  ``Astrophysical implications of the neutrino rest mass. II. The
  density-perturbation spectrum and small-scale fluctuations in the microwave
  background,''
  Sov.\ Astron.\ Lett.\  {\bf 6} (1980) 252
  [Pisma Astron.\ Zh.\  {\bf 6} (1980) 457].


\bibitem{bib:khlopov}
A.~G.~Doroshkevich and M.~Y.~Khlopov,
``The Development of Structure in a Neutrino Universe,''
Sov.\ Astron.\ Lett.\  {\bf 25} (1981) 521.


\bibitem{Shafi:1984ek}
  Q.~Shafi and F.~W.~Stecker,
  ``Implications Of A Class Of Grand Unified Theories For Large Scale Structure
  In The Universe,''
  Phys.\ Rev.\ Lett.\  {\bf 53} (1984) 1292.

\bibitem{Schaefer:1989ua}
  R.~K.~Schaefer, Q.~Shafi and F.~W.~Stecker,
  ``Large scale structure formation and cosmic microwave anisotropy in
a cold plus hot dark matter universe,''
  Astrophys.\ J.\  {\bf 347} (1989) 575.


\bibitem{Lesgourgues:2006nd}
  J.~Lesgourgues and S.~Pastor,
  ``Massive neutrinos and cosmology,''
  Phys.\ Rept.\  {\bf 429} (2006) 307
  [arXiv:astro-ph/0603494].


\bibitem{Hannestad:2006zg}
  S.~Hannestad,
  ``Primordial neutrinos,''
  Ann.\ Rev.\ Nucl.\ Part.\ Sci.\  {\bf 56}, 137 (2006)
  [arXiv:hep-ph/0602058].

\bibitem{Abdalla:2007ut}
  F.~B.~Abdalla and S.~Rawlings,
  ``Determining neutrino properties using future galaxy redshift surveys,''
  Mon.\ Not.\ Roy.\ Astron.\ Soc.\ {\bf 381} (2007) 1313
  [arXiv:astro-ph/0702314].
 
   
\bibitem{Hannestad:2007cp}
  S.~Hannestad and Y.~Y.~Y.~Wong,
  ``Neutrino mass from future high redshift galaxy surveys: Sensitivity and
  detection threshold,''
  JCAP {\bf 0707} (2007) 004
  [arXiv:astro-ph/0703031].

  
\bibitem{Hannestad:2006as}
  S.~Hannestad, H.~Tu and Y.~Y.~Y.~Wong,
  ``Measuring neutrino masses and dark energy with weak lensing tomography,''
  JCAP {\bf 0606} (2006) 025
  [arXiv:astro-ph/0603019].
 
 

\bibitem{Kitching:2008dp}
  T.~D.~Kitching, A.~F.~Heavens, L.~Verde, P.~Serra and A.~Melchiorri,
  ``Finding Evidence for Massive Neutrinos using 3D Weak Lensing,''
  Phys.\ Rev.\  D {\bf 77} (2008) 103008
  [arXiv:0801.4565 [astro-ph]].
  
\bibitem{Lesgourgues:2005yv}
  J.~Lesgourgues, L.~Perotto, S.~Pastor and M.~Piat,
  ``Probing neutrino masses with CMB lensing extraction,''
  Phys.\ Rev.\  D {\bf 73} (2006) 045021
  [arXiv:astro-ph/0511735].
 
  
\bibitem{Gratton:2007tb}
  S.~Gratton, A.~Lewis and G.~Efstathiou,
  ``Prospects for Constraining Neutrino Mass Using Planck and Lyman-Alpha
  Forest Data,''
  Phys.\ Rev.\  D {\bf 77} (2008) 083507
  [arXiv:0705.3100 [astro-ph]].

\bibitem{Ichikawa:2005hi}
  K.~Ichikawa and T.~Takahashi,
  ``On the determination of neutrino masses and dark energy evolution from the
  cross-correlation of CMB and LSS,''
  JCAP {\bf 0802} (2008) 017
  [arXiv:astro-ph/0510849].
  
\bibitem{Lesgourgues:2007ix}
  J.~Lesgourgues, W.~Valkenburg and E.~Gazta{\~n}aga,
  ``Constraining neutrino masses with the ISW-galaxy correlation function,''
  Phys.\ Rev.\  D {\bf 77} (2008) 063505
  [arXiv:0710.5525 [astro-ph]].
    
\bibitem{Pritchard:2008wy}
  J.~R.~Pritchard and E.~Pierpaoli,
  ``Constraining massive neutrinos using cosmological 21 cm observations,''
  arXiv:0805.1920 [astro-ph].
  
\bibitem{Brandbyge:2008rv}
  J.~Brandbyge, S.~Hannestad, T.~Haugb{\o}lle and B.~Thomsen,
  ``The Effect of Thermal Neutrino Motion on the Non-linear Cosmological Matter
  Power Spectrum,''
  JCAP {\bf 0808} (2008) 020
  [arXiv:0802.3700 [astro-ph]].

  
\bibitem{Abazajian:2004zh}
  K.~Abazajian, E.~R.~Switzer, S.~Dodelson, K.~Heitmann and S.~Habib,
   ``The nonlinear cosmological matter power spectrum with massive  neutrinos.
   I: The halo model,''
   Phys.\ Rev.\  D {\bf 71} (2005) 043507
   [arXiv:astro-ph/0411552].


\bibitem{Hannestad:2005bt}
  S.~Hannestad, A.~Ringwald, H.~Tu and Y.~Y.~Y.~Wong,
  ``Is it possible to tell the difference between fermionic and bosonic hot
  dark matter?,''
  JCAP {\bf 0509} (2005) 014
  [arXiv:astro-ph/0507544].
  

\bibitem{Saito:2008bp}
  S.~Saito, M.~Takada and A.~Taruya,
  ``Impact of massive neutrinos on nonlinear matter power spectrum,''
  Phys.\ Rev.\ Lett.\  {\bf 100} (2008) 191301
  [arXiv:0801.0607 [astro-ph]].

\bibitem{bib:juszkie}	
R.~Juszkiewicz,
``On the evolution of cosmological adiabatic perturbations in the weakly non-linear regime,''
Mon.\ Not.\ Roy.\ Astron.\ Soc.\ {\bf 197} (1981) 931.

\bibitem{bib:vishniac}
E.~T.~Vishniac,
``Why weakly non-linear effects are small in a zero-pressure cosmology,''
Mon.\ Not.\ Roy.\ Astron.\ Soc.\ {\bf 203} (1983) 345.

\bibitem{Fry:1983cj}
  J.~N.~Fry,
  ``The Galaxy correlation hierarchy in perturbation theory,''
  Astrophys.\ J.\  {\bf 279} (1984) 499.

\bibitem{Goroff:1986ep}
  M.~H.~Goroff, B.~Grinstein, S.~J.~Rey and M.~B.~Wise,
  ``Coupling of Modes of Cosmological Mass Density Fluctuations,''
  Astrophys.\ J.\  {\bf 311} (1986) 6.


\bibitem{Bernardeau:2001qr}
  F.~Bernardeau, S.~Colombi, E.~Gazta{\~n}aga and R.~Scoccimarro,
  ``Large-scale structure of the universe and cosmological perturbation
  theory,''
  Phys.\ Rept.\  {\bf 367} (2002) 1
  [arXiv:astro-ph/0112551].


\bibitem{Ringwald:2004np}
  A.~Ringwald and Y.~Y.~Y.~Wong,
  ``Gravitational clustering of relic neutrinos and implications for their
  detection,''
  JCAP {\bf 0412} (2004) 005
  [arXiv:hep-ph/0408241].

  
\bibitem{Lewis:1999bs}
  A.~Lewis, A.~Challinor and A.~Lasenby,
  ``Efficient Computation of CMB anisotropies in closed FRW models,''
  Astrophys.\ J.\  {\bf 538} (2000) 473
  [arXiv:astro-ph/9911177].


\bibitem{Ma:1995ey}
  C.~P.~Ma and E.~Bertschinger,
  ``Cosmological perturbation theory in the synchronous and conformal Newtonian
  gauges,''
  Astrophys.\ J.\  {\bf 455} (1995) 7
  [arXiv:astro-ph/9506072].

\bibitem{Hu:1997vi}
  W.~Hu and D.~J.~Eisenstein,
  ``Small scale perturbations in a general MDM cosmology,''
  Astrophys.\ J.\  {\bf 498} (1998) 497
  [arXiv:astro-ph/9710216].
  
  
\bibitem{Fry:1993bj}
  J.~N.~Fry,
  ``The Minimal power spectrum: Higher order contributions,''
  Astrophys.\ J.\  {\bf 421} (1994) 21.


  
\bibitem{Suto:1990wf}
  Y.~Suto and M.~Sasaki,
  ``Quasi Nonlinear Theory Of Cosmological Selfgravitating Systems,''
  Phys.\ Rev.\ Lett.\  {\bf 66} (1991) 264.

\bibitem{Makino:1991rp}
  N.~Makino, M.~Sasaki and Y.~Suto,
  ``Analytic approach to the perturbative expansion of nonlinear gravitational
  fluctuations in cosmological density and velocity fields,''
  Phys.\ Rev.\  D {\bf 46} (1992) 585.
  
\bibitem{Dunkley:2008ie}
  J.~Dunkley {\it et al.}  [WMAP Collaboration],
  ``Five-Year Wilkinson Microwave Anisotropy Probe (WMAP) Observations:
  Likelihoods and Parameters from the WMAP data,''
  arXiv:0803.0586 [astro-ph].
  
\bibitem{Kiakotou:2007pz}
  A.~Kiakotou, {\O}.~Elgar{\o}y and O.~Lahav,
  ``Neutrino Mass, Dark Energy, and the Linear Growth Factor,''
  Phys.\ Rev.\  D {\bf 77} (2008) 063005
  [arXiv:0709.0253 [astro-ph]].

 

\bibitem{Jain:1993jh}
  B.~Jain and E.~Bertschinger,
  ``Second order power spectrum and nonlinear evolution at high redshift,''
  Astrophys.\ J.\  {\bf 431} (1994) 495
  [arXiv:astro-ph/9311070].


\bibitem{Jeong:2006xd}
  D.~Jeong and E.~Komatsu,
  ``Perturbation Theory Reloaded: Analytical Calculation of Non-linearity in
  Baryonic Oscillations in the Real Space Matter Power Spectrum,''
  Astrophys.\ J.\  {\bf 651} (2006) 619
  [arXiv:astro-ph/0604075].

\end{thebibliography}
\end{document}